\documentstyle[aps,prd,epsfig,floats,amsfonts]{revtex}

\newcommand{\beq}{\begin{equation}}
\newcommand{\eeq}{\end{equation}}
\newcommand{\bea}{\begin{eqnarray}} 
\newcommand{\eea}{\end{eqnarray}}
\newcommand{\ba}{\begin{array}}
\newcommand{\ea}{\end{array}}

\begin{document}
\draft
\title{Practical speed meter designs for QND gravitational-wave interferometers}

\author{Patricia Purdue and Yanbei Chen}

\address{
Theoretical Astrophysics, California Institute of Technology,
Pasadena, California 91125
}

\date{\today}
\maketitle
\begin{abstract}

In the quest to develop viable designs for third-generation
optical interferometric gravitational-wave detectors (e.g., LIGO-III
and EURO), one strategy is to monitor the relative 
momentum or speed of the test-mass mirrors, rather than monitoring
their relative position.  A previous paper analyzed a
straightforward but impractical design for a {\it speed-meter
interferometer} that accomplishes this.  This paper describes some
practical variants of speed-meter interferometers.  Like 
the original interferometric speed meter,
these designs {\it in principle} can beat the 
gravitational-wave standard quantum limit (SQL) by 
an arbitrarily large amount, over an arbitrarily wide range
of frequencies.  These variants essentially consist of a 
Michelson interferometer plus an extra ``sloshing" cavity that sends the 
signal back into the interferometer with opposite phase shift, thereby
cancelling the position information and leaving a net phase 
shift proportional
to the relative velocity.  {\it In practice}, the sensitivity of 
these variants will be limited by the maximum light power
$W_{\rm circ}$ circulating in the arm cavities that the mirrors can 
support and by the leakage of vacuum into the optical train at
dissipation points.  In the absence of dissipation and with squeezed
vacuum (power squeeze factor $e^{-2R} \simeq 0.1$) inserted into the
output port so as to keep the circulating power down, the SQL can
be beat by 
$h/h_{\rm SQL} \sim \sqrt{W_{\rm circ}^{SQL} e^{-2R}/W_{\rm circ}}$ 
at all frequencies below some chosen $f_{\rm opt} \simeq 100$ Hz.  Here 
$W_{\rm circ}^{SQL} \simeq 800 {\rm kW} (f_{\rm opt}/100{\rm Hz})^3$ 
is the power required
to reach the SQL in the absence of squeezing.  
[However, as the power increases in this expression,
the speed meter becomes more narrow band; additional
power and re-optimization of some parameters are required to maintain the
wide band.  See Sec.~\ref{sec:optimization}.]
Estimates are given of 
the amount by which vacuum
leakage at dissipation points will debilitate this sensitivity 
(see Fig.~\ref{fig:lossynoise}); these losses are 10\% or less over
most of the frequency range of interest ($f\gtrsim 10~\rm Hz$).  The
sensitivity can be improved, particularly at high freqencies,
by using frequency-dependent homodyne detection, which unfortunately
requires two 4-kilometer-long filter cavities (see Fig.~\ref{fig:threearmfull}).

\end{abstract}
\pacs{PACS numbers: 04.80.Nn, 95.55.Ym, 42.50.Dv, 03.67.-a}

\narrowtext
\twocolumn

\section{Introduction}
\label{sec:Introduction}

This paper is part of the effort to explore theoretically various 
ideas for a third-generation interferometric gravitational-wave 
detector.  The goal of such detectors is to beat, by a factor of 5 
or more, the {\it standard quantum limit} (SQL)---a limit that
constrains interferometers \cite{CavesSQLforIFOs} such as 
LIGO-I which have conventional 
optical topology \cite{ligoIa,ligoIb}, but does not constrain more 
sophisticated ``quantum nondemolition''
(QND) interferometers \cite{Unruhsqueezedvacuum,JR}.

The concepts currently being explored for third-generation detectors
fall into two categories:  {\it external readout} and {\it intracavity
readout}.  In interferometer designs with external readout 
topologies, light exiting the interferometer is monitored for phase
shifts, which indicate the motion of the test masses.  Examples 
include conventional interferometers and their variants 
(such as LIGO-I \cite{ligoIa,ligoIb},
LIGO-II \cite{ligoII}, and those discussed 
in Ref.~\cite{Kimble}), as well as the 
speed-meter interferometers discussed here and in a previous paper \cite{Purdue}.
In intracavity readout topologies, the gravitational-wave
force is fed via light pressure onto a tiny internal mass, whose displacement
is monitored with a local position transducer.
Examples include the optical bar, symphotonic
state, and optical lever schemes discussed by Braginsky, Khalili, and Gorodetsky
\cite{opticalbar,symphotonicstates,opticallever}.  These
intracavity readout interferometers may be able to function at
much lower light powers than external readout interferometers of comparable
sensitivity because the QND readout is performed via the local position
transducer (perhaps microwave-technology based), instead of via the 
interferometer's light; however, the designs are not yet fully
developed.

At present, the most complete analysis of candidate designs for 
third-generation external-readout detectors has been carried out by 
Kimble, Levin, Matsko, Thorne, and Vyatchanin \cite{Kimble} (KLMTV). 
They examined three potential designs for interferometers that could 
beat the SQL: a squeezed-input interferometer, which makes use of
squeezed vacuum being injected into the dark port; a variational-output
scheme in which frequency-dependent homodyne detection was used; and
a squeezed-variational interferometer that combines the features of
both.  (Because the KLMTV designs measure the relative positions
of the test masses, we shall refer to them as {\it position meters}, particularly
when we want to distinguish them from the speed meters that, for example,
use variational-output techniques.)  Although at least some of the KLMTV
position-meter designs have remarkable performance in the lossless limit, 
all of them are highly susceptible
to losses. 

In addition, we note that the KLMTV position meters each require four
kilometer-scale cavities (two arm cavities + two filter cavities).  The 
speed meters described in this paper require at least three kilometer-scale
cavities [two arm cavities + one ``sloshing" cavity (described below)].
If we use a variational-output technique, as KLMTV did, the resulting
interferometer will have five kilometer-scale cavities (two arm cavities + 
one sloshing cavity + two filter cavities).  This is shown in 
Fig.~\ref{fig:threearmfull} below.

\begin{figure}[t]
\epsfig{file=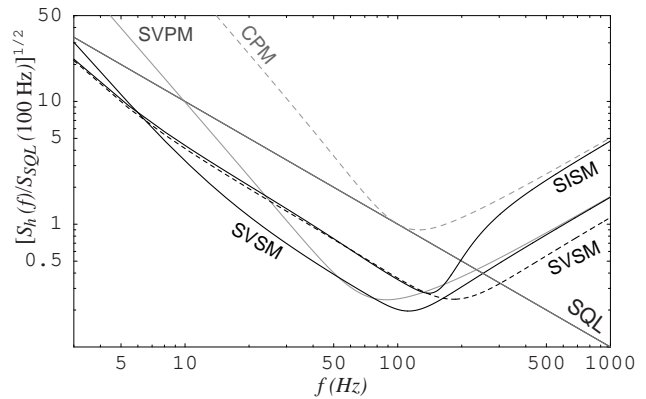,width=0.475\textwidth} 
\caption[Comparison of typical noise curves
(with losses) for several interferometer configurations]
{Comparison of noise curves (with losses) of
several interferometer configurations.  Each of these
curves has been optimized in a way that is meant to
illustrate their relative advantages and disadvantages.
The conventional position meter (CPM) \protect\cite{Kimble} 
has $W_{\rm circ}=820~\rm kW$ and 
bandwidth $\gamma=cT/4L=2\pi \times 100~\rm Hz$.  The squeezed-input
speed meter (SISM)---optmized to
agree with the conventional position meter at high 
frequencies---has power squeeze factor $e^{-2R}=0.1$, 
optimal frequency $\omega_{\rm opt}=2\pi \times 105~\rm Hz$, 
extraction rate $\delta=2\omega_{\rm opt}$, and
sloshing frequency $\Omega=\sqrt{3}\omega_{\rm opt}$.  
The squeezed-variational position meter (SVPM) \protect\cite{Kimble}
has the same parameters as the conventional position meter,
with power squeeze factor $e^{-2R}=0.1$.
There are two squeezed-variational speed-meter curves (SVSM).  
One (black dashes) uses the
same parameters as the squeezed-input speed meter.  The other (solid curve)
has been optimized to compare more directly with the
squeezed-variational position meter; it has
$\Omega=2\pi \times 95~\rm Hz$ and 
$\delta = 2\pi \times 100~\rm Hz$ (note that our $\delta$ is
equivalent to the bandwidth $\gamma$ used to describe
the interferometers in Ref.~\protect\cite{Kimble}).}
\label{fig:introcompare}
\end{figure}

The speed meter described in this paper can achieve a performance 
significantly better than a conventional position meter, as shown in
Fig.~\ref{fig:introcompare}. (By ``conventional," we mean ``without 
any QND techniques." An example is LIGO-I.)  The squeezed-input
speed meter (SISM) noise curve shown in 
that Fig.~\ref{fig:introcompare} beats the SQL by a factor
of $\sqrt{10}$ in amplitude 
and has {\it fixed-angle} squeezed vacuum injected into the dark port
[this allows the interferometer to operate at a lower circulating power
than would otherwise be necessary to achieve that level of sensitivity, 
as described by Eq.~(\ref{hratio}) below].
The squeezed-variational position meter (SVPM), which requires
{\it frequency-dependent} squeezed vacuum and homodyne detection,
is more sensitive than the squeezed-input speed meter over much of the frequency
range of interest, but the speed meter has the 
advantage at low frequencies.  It should also be noted that 
the squeezed-variational position meter requires four kilometer-scale
cavities (as described in the previous paragraph), whereas the
squeezed-input speed meter requires three.

If frequency-dependent homodyne
detection is added to the squeezed-input speed meter, the resulting 
squeezed-variational speed meter (SVSM) can be 
optimized to beat the 
squeezed-variational position meter over the entire frequency range.  Figure~\ref{fig:introcompare}
contains two squeezed-variational speed meter curves; one is optimized
to match the squeezed-input speed meter curve at low frequencies, and the
other is optimized for comparison with the squeezed-variational postion-meter curve
(resulting in less sensitivity at high frequencies).

The original idea for a speed meter, as a device for measuring the 
momentum of a single test mass, was conceived by Braginsky and Khalili
\cite{firstspeed} and was further developed by Braginsky, Gorodetsky, Khalili, and 
Thorne \cite{Brag} (BGKT).  In their appendix, BGKT sketched a design for 
an interferometric gravity wave speed meter and speculated that it would
be able to beat the SQL.  This was verified in Ref.~\cite{Purdue} 
%(which appears
%as Chapter 2 of this thesis),
(Paper~I),
%\footnote{Paper~I (Ref.~\cite{Purdue}) appears as Chapter 
%\ref{chap:firstspeed} in this thesis.},
where it was
demonstrated that such a device could {\it in principle}
beat the SQL by an arbitrary amount over a wide range of frequencies.  
However, the design presented in that paper, which we shall call
the {\it two-cavity speed-meter} design, had three
significant problems: it required (i) a high circulating power
($\sim$ 8 MW to beat the SQL by a factor of 10 in noise power at 100 Hz and below),
(ii) a large amount of power coming out of the interferometer
with the signal ($\sim$ 0.5 MW), and (iii) an exorbitantly 
high input laser power ($\gtrsim$ 300 MW).
The present paper describes an alternate class 
of speed meters that effectively eliminate the latter two problems, and
techniques for reducing the needed circulating power are discussed.  These
improvements bring 
interferometric speed meters into the realm of practicality.

\begin{figure}
\epsfig{file=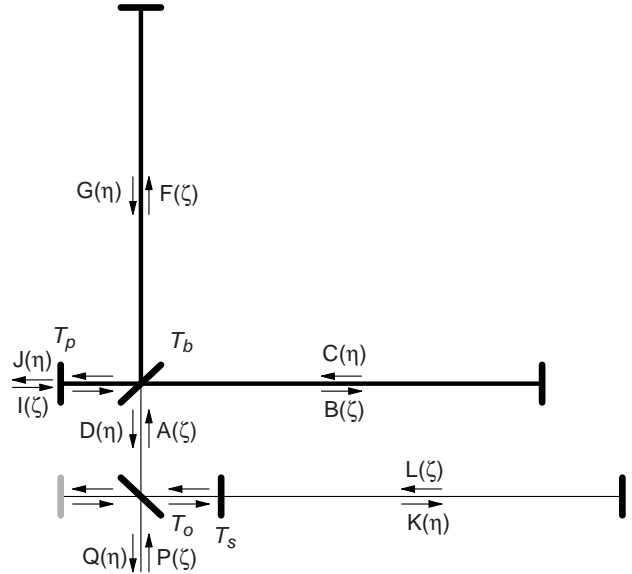,width=0.475\textwidth} 
\caption[Simple version of three-cavity design for speed-meter interferometer]
{Simple version of three-cavity design for speed-meter interferometer.  
The main laser input port is denoted by $I(\zeta)$, where $\zeta=t-z/c$.  
The signal is extracted at the bottom mirror [denoted $Q(\eta)$, where 
$\eta=t+z/c$].  The difference
between the one- and two-port versions is the mirror shown in gray.}
\label{fig:oneport}
\end{figure}

A simple version of the {\it three-cavity speed-meter} design to be 
discussed in this paper is shown in Fig.~\ref{fig:oneport}.
In (an idealized theorist's version of) this speed meter, the input laser light 
[with electric field denoted $I(\zeta)$ in Fig.~\ref{fig:oneport}] passes 
through a power-recycling mirror into a standard Michelson interferometer.  
The relative phase shifts of the two arms are adjusted
so that all of the input light returns to the input port, leaving the
other port dark [i.e., the interferometer is operating in the
symmetric mode so $D(\eta)=0$ in Fig.~\ref{fig:oneport}].  
In effect, we have a resonant cavity shaped
like $\perp$.  When the end mirrors move, they will
put a phase shift on the light, causing some light to enter
the antisymmetric mode (shaped like $\vdash$) and come out the dark port.
So far, this is the same as conventional interferometer designs 
(but without the optical cavities in the two interferometer arms).  

Next, we feed the light coming out of the dark port [$D(\eta)$] 
into a sloshing cavity
[labeled $K(\eta)$ and $L(\zeta)$ in Fig.~\ref{fig:oneport}].  The light
carrying the position information sloshes back into the ``antisymmetric cavity"
with a phase shift of $180^\circ$, cancelling the position 
information in that cavity and leaving only a phase shift proportional
to the relative velocity of the test masses\footnote{The net signal is 
proportional to the relative velocities of the test masses, 
assuming that the frequencies $\omega$ of the test masses' motion 
are $\omega \ll \Omega = (\rm sloshing\ frequency)$.
However, the optimal regime of operation for the 
speed meter is $\omega \sim \Omega$.  As a result, the output signal
contains a sum over odd time derivatives of position
[see the discussion in Sec.~\ref{sec:newlosslessmath}].  
Therefore, the speed meter monitors not just the
relative speed of the test masses, but a mixture of
all odd time derivatives of the relative positions of the test masses.}.
The sloshing frequency is
\begin{equation}
\Omega = \frac{c\sqrt{T_{\rm s}}}{2L} \,,
\label{sloshdef}
\end{equation}
where $T_{\rm s}$ is the power transmissivity of the sloshing mirror, 
$L$ is the common length of all three cavities, and $c$ is the speed of 
light.
We read the velocity signal [$Q(\eta)$] 
out at a extraction mirror (with transmissivity $T_{\rm o}$), which gives 
a signal-light extraction rate of
\begin{equation}
\delta = \frac{c T_{\rm o}}{L} \,.
\label{extractdef}
\end{equation}
We have used the extraction mirror to put the sloshing cavity parallel to
one of the arms of the Michelson part of the interferometer, allowing this
interferometer to fit into the existing LIGO facilities.  The presence
of the extraction mirror essentially opens two ports to our system.  We can
use both outputs, or we can add an additional mirror to close one port
(the gray mirror in Fig.~\ref{fig:oneport}).  
We will focus on the latter case in this paper.
 
The sensitivity $h$ of this
interferometer, compared to the SQL, can be expressed as\footnote{It 
should be noted that, as the power increases in Eq.~(\ref{hratio}),
the speed-meter performance becomes more narrow band.  Additional
power and a re-optimization of some of the speed meter's parameters are 
required to maintain the same bandwidth at higher sensitivities. 
See Sec.~\ref{sec:optimization} for details.}
\beq
\frac{h}{h_{SQL}} \sim \sqrt{\frac{W_{\rm circ}^{SQL}}{ e^{2R} W_{\rm circ}}}
\simeq \sqrt{\frac{800~\rm kW}{e^{2R} W_{\rm circ}}} \,,
\label{hratio}
\eeq
where $W_{\rm circ}$ is the power circulating in the arms,
$W_{\rm circ}^{SQL} \simeq 800 {\rm kW} (f_{\rm opt}/100{\rm Hz})^3$ 
is the power required
to reach the SQL in the absence of squeezing
(for the arms of 
length $L=4~\rm km$ and test masses with mass $m=40~\rm kg$),
and $e^{2R}$ is the power
squeeze
factor\footnote{For an explanation of squeezed vacuum and squeeze factors,
see, for example, KLMTV and references cited therein.  In particular, 
their work was
based on that of Caves \cite{CavesSqVac} and Unruh \cite{Unruhsqueezedvacuum}.
Also, KLMTV state that a likely achievable value for the squeeze factor
(in the LIGO-III time frame) is $e^{2R} \simeq 10$, so we use that value
in our discussion.}.
%and more recently by
%Jaekel and Reynaud \cite{JR} and Pace {\it et al.} \cite{PCW}.} 
With no squeezed vacuum, the squeeze factor is $e^{2R} = 1$, so
the circulating power $W_{\rm circ}$ must be $8~\rm MW$
in order to beat the SQL at $f_{\rm opt} \simeq 100~\rm Hz$ by a 
factor of $\sqrt{10}$ in sensitivity. 
With a squeeze factor of $e^{2R} = 10$,
we can achieve the same performance with 
$W_{\rm circ} \simeq 800~\rm kW$, which is the
same as LIGO-II is expected to be.

\begin{figure}[t]
\epsfig{file=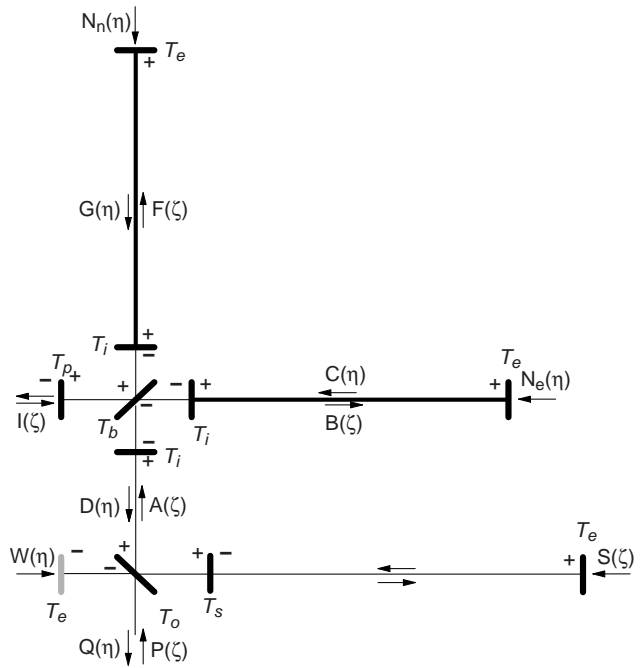,width=0.475\textwidth} 
\caption[Practical three-cavity speed-meter design]
{Schematic diagram showing the practical version of the three-cavity 
speed-meter design, which reduces
the power flowing through the beam splitter.  Three additional mirrors,
with transmissivity $T_{\rm i}$, are placed around the beam splitter.   
The ``$+$" and ``$-$" 
signs near the mirrors indicate the sign of
the reflectivities in the junction conditions for each location.  
The mirror shown in gray closes the second port of the interferometer.}
\label{fig:oneport3IM}
\end{figure}

This performance (in the lossless limit) 
is the same as that of the  
two-cavity (
%Chapter 2)
Paper~I) 
speed meter for the same 
circulating power, but the three-cavity design has an overwhelming 
advantage in terms of required input power. 
However, there is one 
significant problem with 
this design 
that we must address: the uncomfortably large amount of laser power,
equal to $W_{\rm circ}$, 
flowing through the beam splitter.  Even with the use of squeezed 
vacuum, this power will be too high. 
Fortunately, there is a method, based on the work of Mizuno \cite{M95},
that will let us solve this problem:

We add three mirrors into our speed meter (labeled
$T_{\rm i}$ in Fig.\ \ref{fig:oneport3IM});
we shall call this the {\it practical three-cavity speed meter}. 
Two of the additional mirrors are placed in the excited arms of the
interferometer to create resonating Fabry-Perot
cavities in each arm (as for conventional interferometers such as LIGO-I). 
The third mirror is added between the beam
splitter and the extraction mirror, in such a way that 
light with the carrier frequency 
resonates in the subcavity formed by this mirror and the internal mirrors.

As claimed by Mizuno \cite{M95} and tested experimentally by Freise et
al.~\cite{FHSMSLWSRWD00} and Mason \cite{M01}, when the transmissivity
of the third mirror decreases from $1$, the storage time of sideband
fields in the arm cavity due to the presence of the internal
mirrors will decrease.  This phenomenon is called Resonant Sideband
Extraction (RSE); consequently, the third mirror is called the 
RSE mirror.  One special case, which is of great 
interest to us, occurs when
the RSE mirror has the same transmissivity as the
internal mirrors.  In this case, the effect of the internal
mirrors on the gravitational-wave sidebands should
be exactly cancelled out by the RSE mirror.  The three new mirrors 
then have just one effect: they reduce
the carrier power passing through the beam splitter---and they
can do so by a large factor.

Indeed, we have confirmed that this is true for our speed
meter, as long as the distances between the three additional mirrors 
(the length of the ``RSE cavity'') are
small (a few meters), so that the phase shifts added to the 
slightly off-resonance sidebands by the RSE cavity are negligible.  
We can then adjust the transmissivities of the power-recycling mirror and of 
the three internal mirrors to reduce the amount of carrier power passing 
through the beam splitter to a more reasonable level. 

With this design, the high circulating power is confined to the Fabry-Perot
arm cavities, as in conventional LIGO designs.  There is some question
as to the level of power that mirrors will be able to tolerate in the
LIGO-III time frame.  Assuming that several megawatts is not acceptable,
we shall show that the circulating power can be reduced by injecting
fixed-angle squeezed vacuum into the dark port, as indicated by
Eq.~(\ref{hratio}).  

Going a step farther, we shall
show that if, in addition to injected squeezed
vacuum, we also use frequency-dependent (FD) homodyne detection, the sensitivity
of the speed meter is dramatically improved at high frequencies (above 
$f_{\rm opt} \simeq 100$ Hz); 
this is shown in Fig.~\ref{fig:introcompare}.
%See Sec.~\ref{sec:sqvacFDdetect}.  
The disadvantage of this is that 
FD homodyne detection requires two filter
cavities of the same length as the arm cavities (4 km for LIGO), as shown in
Fig.~\ref{fig:threearmfull}.

\begin{figure}
\epsfig{file=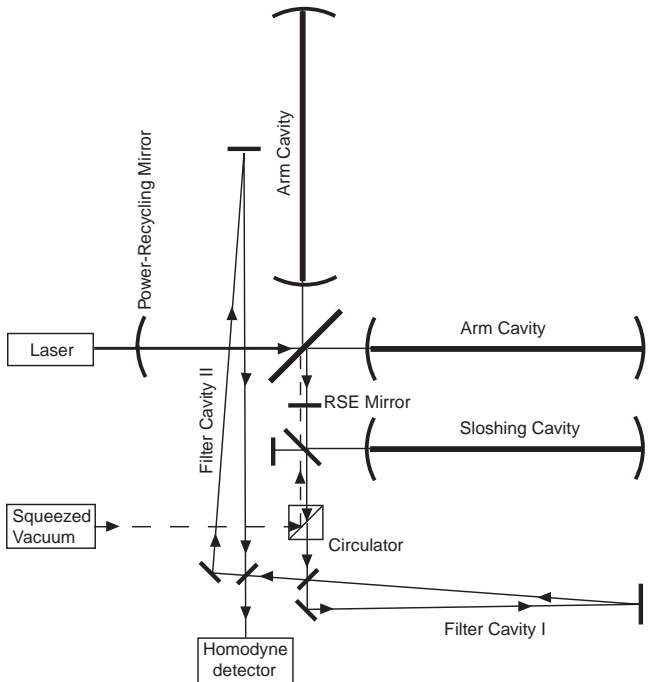,width=0.475\textwidth} 
\caption[Squeezed-variational three-cavity speed-meter design]
{Schematic diagram showing the practical three-cavity speed-meter design with 
squeezed vacuum injected at the dark port and two filter cavities on the
output.  Note that the circulator is a four-port optical device that separates
the injected (squeezed) input and the interferometer's output.}
\label{fig:threearmfull}
\end{figure}

Our analysis of the losses in these scenarios indicates that our speed meters
with squeezed vacuum and/or variational-output are much less sensistive to
losses than
a position meter using those techniques (as analyzed by KLMTV).
Losses
for the various speed meters we discuss here are generally quite low
and are due primarily to the losses in the optical elements (as opposed
to mode-mismatching effects). Without
squeezed vacuum, the losses in sensitivity are less than 10\% in the range 
$50-105~\rm Hz$, lower at higher frequencies, but higher at low frequencies.
Injecting fixed-angle squeezed vacuum into the dark port allows 
this speed meter to operate
at a lower power [see Eq.~\ref{hratio}], thereby reducing the 
dominant losses (which are dependent on the circulating power because they come
from vacuum fluctations contributing to the back-action).  In this case, the losses are less
than 4\% in the range $25-150~\rm Hz$.  As before, they are lower
at high frequencies, but they increase at low frequencies.
Using FD homodyne detection does not change the losses significantly.

This paper is organized as follows:  In Sec.~\ref{sec:shortmath} we give a 
brief description of the mathematical method that we use to analyze the 
interferometer.  In Sec.~\ref{sec:newlosslessmath}, we present the 
results in the lossless case,
followed in Sec.~\ref{sec:optimization} by a discussion of optimization methods.
In Sec.~\ref{sec:basicdisc}, we discuss some of the 
advantages and disadvantages of this design, including
the reasons it requires a large circulating power.
Then in Sec.~\ref{sec:modifications}, we show how the circulating power can be 
reduced by injecting squeezed vacuum through the dark port of the interferometer
and how the use of frequency-dependent homodyne detection can improve the
performance at high frequencies.
In Section~\ref{sec:newlossy}, we discuss
the effect of losses on our speed meter with the various modifications
made in Sec.~\ref{sec:modifications}, and we compare our interferometer
configurations with those of KLMTV.  Finally, we summarize our results 
in Sec.~\ref{sec:conclusions}.

\section{Mathematical Description of the Interferometer}
\label{sec:shortmath}

The interferometers in this paper are analyzed using the techniques
described in %Sec.~\ref{sec:math}.
Paper~I (Sec.~II).   
These methods are 
based on the formalism developed by Caves and Schumaker \cite{cs,sc}
and used by KLMTV
 to examine more conventional interferometer designs.
For completeness, we will summarize the main points here.

The electric field propagating in each direction down each
segment of the interferometer is expressed in the form
\begin{equation}
E_{\rm field}(\zeta) = \sqrt{\frac{4 \pi \hbar \omega_0}{{\cal S} c}} A(\zeta) \,.
\end{equation}
Here $A(\zeta)$ is the amplitude (which is denoted by other 
letters---$B(\zeta)$, $P(\zeta)$, etc.---in other parts of the 
interferometer; see Fig.\ \ref{fig:oneport}), 
$\zeta = t - z/c$, $\omega_0$ is the carrier
frequency, $\hbar$ is the reduced Planck's constant, 
and ${\cal S}$ is the effective cross-sectional area of the light
beam; see Eq.~(8) of KLMTV.  For light propagating in the negative
$z$ direction, $\zeta = t - z/c$ is replaced by $\eta = t + z/c$.  
We decompose the amplitude into cosine and
sine quadratures,
\begin{equation}
A(\zeta) = {\cal A}_1(\zeta) \cos \omega_0 \zeta 
+ {\cal A}_2(\zeta) \sin \omega_0 \zeta \,,
\end{equation}
where the subscript 1 always refers to the cosine 
quadrature, and 2 to sine.  Both arms and the sloshing cavity have length $L=4$ km,
whereas all of the other lengths $z_i$ are short compared to $L$.
We choose the cavity lengths to be exact half multiples of the
carrier wavelength so $e^{i 2 \omega_0 L/c} =1$ and $e^{i 2 \omega_0 z_i/c} =1$.  
There will be phase shifts put onto the sideband light in all of these
cavities, but only the
phase shifts due to the long cavities are significant.

The aforementioned sidebands are put onto the carrier by the mirror motions and by
vacuum fluctuations.  We express the quadrature
amplitudes for the carrier plus the side bands in the form
\begin{equation}
{\cal A}_j(\zeta) =
A_j(\zeta) + \int_0^\infty \bigl[ \tilde{a}_j(\omega) e^{-i \omega \zeta}
+ \tilde{a}_j^{\dagger}(\omega) e^{i \omega \zeta}\bigr]
\frac{d\omega}{2\pi} \,.
\label{sidebandeq}
\end{equation}
Here $A_j(\zeta)$ is the carrier amplitude,
$\tilde{a}_j(\omega)$ is the field amplitude (a quantum mechanical
operator) for the sideband
at sideband frequency $\omega$ (absolute frequency $\omega_0 \pm \omega$)
in the $j$ quadrature, and $\tilde{a}_j^\dagger (\omega)$ is the Hermitian
adjoint of $\tilde{a}_j(\omega)$; cf.\ Eqs.\ (6)--(8) of KLMTV,
where commutation relations and the connection to creation and annihilation
operators are discussed.  In other portions of the interferometer 
(Fig.~\ref{fig:oneport}), ${\cal A}_j(\zeta)$ is replaced by, e.g.,
${\cal C}_j(\zeta)$; $A_j(\zeta)$, by $C_j(\zeta)$; $\tilde{a}_j(\omega)$,
by $\tilde{c}_j(\omega)$, etc.

Since each mirror has a power transmissivity and complementary reflectivity 
satisfying the equation $T+R=1$, we can write out the junction conditions
for each mirror in the system, for both the carrier quadratures and
the sidebands 
%[see particularly Eqs.~(\ref{eqs:junction}) and 
%(\ref{eqs:Aeq})--(\ref{eqs:Ceq})].
[see particularly Eqs.\ (5) and (12)--(14) in Paper~I].
We shall denote the power transmissivities for the sloshing 
mirror as $T_{\rm s}$, for the extraction (output) mirror as $T_{\rm o}$, 
the power-recycling mirror as $T_{\rm p}$, for the beam-splitter as $T_{\rm b}=0.5$, 
for the internal mirrors as $T_{\rm i}$,
and for the end mirrors as $T_{\rm e}$; see Figs.~\ref{fig:oneport}
and \ref{fig:oneport3IM}.  

The resulting equations can be solved simultaneously to get
expressions for the carrier and sidebands in each segment of
the interferometer.  Since those expressions may be quite complicated,
we use the following assumptions to simplify our results.  First, we assume
that only the cosine quadrature is being driven (so that the carrier
sine quadrature terms are all zero).  Second, we assume that the transmissivities
obey 
\begin{equation}
1 \gg T_{\rm o} \gg T_{\rm s} \gg T_{\rm e} \quad {\rm and} \quad 
1 \gg \{ T_{\rm p}, T_{\rm i} \} \gg T_{\rm e} \,.
\label{newtransreq}
\end{equation}
The motivations for these assumptions are that (i) they lead to speed-meter 
behavior; (ii) as with any interferometer, the best performance is achieved by
making the end-mirror transmissivities $T_{\rm e}$ as small as possible;
and (iii) good performance requires a light extraction rate 
comparable to the sloshing
rate, $\delta \sim \Omega$ [cf.\ the first paragraph of 
Sec.~III B in Paper~I],
%Sec.~\ref{sec:losslessnum}], 
which with
Eqs.\ (\ref{sloshdef}) and (\ref{extractdef}) implies 
$T_{\rm o} \sim \sqrt{T_{\rm s}}$
so $T_{\rm o} \gg T_{\rm s}$.  Throughout the paper, we will be
using these assumptions, together with $\omega L/c \ll 1$, to simplify
our expressions.

%This procedure yields an output [$Q(\eta)$] containing an
%$\omega (\tilde{x}_e-\tilde{x}_n)$ term, where $\tilde{x}_e$ is the
%Fourier transform of the displacement of the ``east" mirror and 
%$\tilde{x}_n$ is the same for the ``north" mirror.  The $\omega %(\tilde{x}_e-\tilde{x}_n)$ is the Fourier 
%transform of the end-mirror
%velocity (relative to the corresponding internal mirror).  Since
%there is no factor of $\tilde{x}_e$ or $\tilde{x}_n$ without 
%a multiplying factor $\omega$ in the
%output, our interferometer is indeed a speed meter.

\section{Speed Meter in the Lossless Limit}
\label{sec:lossless}

For simplicity, in this section we will set $T_{\rm e}=0$ (end
mirrors perfectly reflecting).  We will
also neglect the (vacuum-fluctuation) noise coming 
in the main laser port ($\tilde{i}_{1,2}$)
since that noise largely exits back toward the laser and produces 
negligible noise on the signal light exiting the output port.
As a result of these assumptions, the only (vacuum-fluctuation) noise
that remains is that which comes in through the output port
($\tilde{p}_{1,2}$). An interferometer in which this is the case
and in which light absorption and scattering are unimportant ($R+T=1$
for all mirrors, as we have assumed) is said to be ``lossless."
In Sec.\ \ref{sec:newlossy}, we shall relax these assumptions; i.e., we
shall consider lossy interferometers.

It should be noted that the results and discussion in this section and
in Sec.~\ref{sec:modifications} apply to both the simple and practical
versions of the three-cavity speed meter (Figs.~\ref{fig:oneport} and
\ref{fig:oneport3IM}).  The two versions are completely
equivalent (in the lossless limit).

\subsection{Mathematical Analysis}
\label{sec:newlosslessmath}

The lossless interferometer output for the speed meters in Fig.~\ref{fig:oneport}
and \ref{fig:oneport3IM},
as derived by the analysis sketched in the previous section, is then
\begin{mathletters}
\begin{eqnarray}
\tilde q_1 &=& - {{\cal L}^*(\omega)\over{\cal L}(\omega)}\tilde{p}_1 \,,
\label{q1out} \\
\tilde{q}_2 &=&
   \frac{2 i \omega \sqrt{\omega_0 \delta W_{\rm circ}}}{\sqrt{\hbar c L}
    {\cal L}(\omega)} \tilde{x}
   -\frac{{\cal L}^\ast (\omega)}{{\cal L}(\omega)} \tilde{p}_2\,.
\label{q2out}
\end{eqnarray}
\label{qsout}
\end{mathletters}
Here $\tilde{p}_j (\omega)$ is the side-band field operator [analog of 
$\tilde{a}_j (\omega)$ in Eq.~(\ref{sidebandeq})] associated with the 
dark-port input $P(\zeta)$, and $\tilde{q}_j (\omega)$ associated with the 
output $Q(\eta)$; see Fig.~\ref{fig:oneport}.  Also, in Eqs.~(\ref{qsout}),
${\cal L}(\omega)$ is a c-number given by
\begin{equation}
{\cal L}(\omega) = \Omega^2 - \omega^2 - i \omega \delta 
\label{fancyL}
\end{equation}
[recalling that $\Omega = c\sqrt{T_{\rm s}}/2L$ is the sloshing frequency,
$\delta = cT_{\rm o}/L$ the extraction rate], 
the asterisk in ${\cal L}^\ast (\omega)$ denotes the
complex conjugate, $\tilde{x}(\omega)$ is the Fourier transform of the 
relative displacement of the four test masses---i.e., the Fourier
transform of the difference in lengths of the interferometer's two arm
cavities---and $W_{\rm circ}$ is the circulating power in the
each of the interferometer's two arms.  Note that the circulating power 
(derived as in 
%Sec.~\ref{sec:losslessside})
Sec.~II B of Paper~I)
is related to the carrier amplitude $B_1$ in the arms 
by\footnote{Equation~(\ref{Wcirc}) refers specifically to the practical
version of the three-arm interferometer (Fig.~\ref{fig:oneport3IM}).  The
simple (Fig.~\ref{fig:oneport}) version would be 
$$ W_{\rm circ}=\frac{1}{2} \hbar \omega_0 B_1^2 
= \frac{\hbar \omega_0 I_1^2}{T_{\rm p}} \,.$$}
\beq
W_{\rm circ}=\frac{1}{2} \hbar \omega_0 B_1^2 
= \frac{4 \hbar \omega_0 I_1^2}{T_{\rm i} T_{\rm p}} \,,
\label{Wcirc}
\eeq
where $I_1$ is the input laser amplitude (in the cosine quadrature).
Readers who wish to derive the input--output relations (\ref{qsout})
for themselves may find useful guidance in Appendix~B of KLMTV \cite{Kimble} and
in 
%Secs.~\ref{sec:math} and \ref{sec:losslessspeed},
Secs.~II and III of Paper~I \cite{Purdue}, 
which give detailed
derivations for other interferometer designs.

Notice that the first term in Eq.~(\ref{q2out}) contains $\tilde{x}$
only in the form $\omega \tilde{x}$; this is the velocity signal
[actually, the sum of the velocity and higher odd time derivatives
of position because of the ${\cal L}(\omega)$ in the denominator].
The test masses' relative displacement $\tilde{x}(\omega)$ 
is given by
\beq
\tilde{x}=\tilde{x}_e -\tilde{x}_n= L\tilde{h} 
- \frac{8 i \sqrt{\hbar \omega_0 \delta W_{\rm circ}} }
{m \omega \sqrt{cL} {\cal L}(\omega)} \tilde{p}_1 \,,
\label{xwba}
\eeq
where $\tilde{x}_e$ is the
Fourier transform of the relative displacement of the mirrors of the 
``east" arm and 
$\tilde{x}_n$ is the same for the ``north" arm.  The last term
is the back-action produced by fluctuating radiation pressure
%(derived as in Sec.~\ref{sec:losslessside}).
(derived as in Sec.~II B of Paper~I).

%With back-action (derived according to the method given in Paper~I),
%the position is given by 
%\begin{equation}
%\tilde{x}_{e,n} = \chi_{en} \biggl( \frac{1}{2} L\tilde{h} 
%- \frac{4 i \sqrt{\hbar \omega_0 T_{\rm o} W_{\rm circ}} }
%{m \omega L {\cal L}(\omega)} \tilde{p}_1 \biggr)\,,
%\label{xwba}
%\end{equation}
%where $\chi_{en} =+1$ for the east arm and $-1$ for the north arm. 

It is possible to express Eqs.~(\ref{qsout}) in a more concise form,
similar to Eqs.\ (16) in KLMTV:
\begin{mathletters}
\begin{eqnarray}
\tilde{q}_1 &=& \Delta \tilde{p}_1= \tilde{p}_1 e^{2i \psi} \,, \\
\tilde{q}_2 &=& \Delta \tilde{p}_2
	+ \sqrt{2 \kappa} \frac{\tilde{h}}{h_{SQL}} e^{i \psi} \,, \quad 
	 \Delta \tilde{p}_2 = (\tilde{p}_2-\kappa \tilde{p}_1) e^{2i \psi} \,.
\end{eqnarray}
\label{qsoutKimbleform}
\end{mathletters}
Here
\beq
\tan \psi = - \frac{\Omega^2-\omega^2}{\omega \delta} 
\label{tanpsi} 
\eeq
is a phase shift put onto the light by the interferometer,
\beq
\kappa
	= \frac{16 \omega_0 \delta W_{\rm circ}}{m c L |{\cal L}(\omega)|^2} 
\label{kappa}
\eeq
is a dimensionless coupling constant that couples the gravity wave signal
$\tilde{h}$ into the
output $\tilde{q}_2$,
and
\begin{equation}
h_{SQL} = \sqrt{\frac{8 \hbar}{m \omega^2 L^2}} 
\end{equation}
is the standard quantum
limit for a conventional interferometer such as LIGO-I or VIRGO 
\cite{CavesSQLforIFOs}. 

\begin{figure}
\epsfig{file=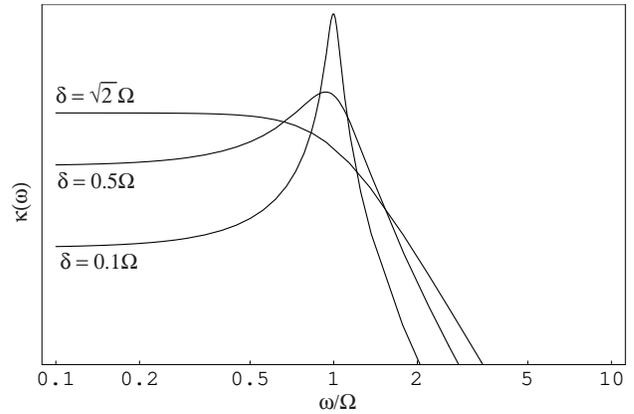,width=0.475\textwidth} 
\caption[Typical curves for the coupling constant $\kappa(\omega)$ in terms
of the sloshing frequency $\Omega$]
{The coupling constant $\kappa(\omega)$ in arbitrary
(logrithmic) units with $\omega$  
measured in units of $\Omega$. The three curves correspond to the same
light power (such that $\kappa_{\rm max}=5$ for the middle curve), but
$\delta=0.1\Omega$, $\delta=0.5\Omega$, and $\delta =\sqrt{2}\Omega$.}
\label{fig:kappafig}
\end{figure}

In Fig.~\ref{fig:kappafig}, we plot the coupling constant 
$\kappa$ as a function of frequency 
for several values of $\delta$. As the graph shows, $\kappa$ can be
roughly constant for a rather broad frequency 
band $\omega \stackrel{<}{_\sim} \Omega$, when
$\delta$ is chosen to be $\sim \Omega$ (as it will be when the interferometer
is optimized).  Combining this with the fact that
$h_{SQL}\propto 1/\omega$, we infer from Eqs.~(\ref{qsoutKimbleform}) 
that the output {\it signal} at frequencies $\omega \lesssim \Omega$
is proportional to $\omega \tilde{h}$, or equivalently $\omega \tilde{x}$, which 
is the relative speed of the test masses (as mentioned above). 

The terms $ \Delta \tilde{p}_1$ and $ \Delta \tilde{p}_2$ in Eqs.~(\ref{qsout})
represent {\it quantum noise} (shot noise, radiation-pressure noise, 
and their correlations). We shall demonstrate below that, in the 
frequency band $\omega \lesssim \Omega$ where
the interferometer samples only the speed, there is no back-action (radiation-pressure) noise. 
This might not be obvious from Eqs.~(\ref{qsoutKimbleform}),
especially because they have an identical form (except for the frequency
dependence of $\kappa$) as the input-output
relations of a conventional interferometer, where the term 
proportional to $\cal K$ (their version of $\kappa$) 
{\it is} the radiation-pressure noise. Indeed, if one measures the ``sine"
quadrature of the output, $\tilde{q}_2$, as is done in a conventional
interferometer, this speed meter turns out to be SQL limited, as 
are conventional interferometers. 

Fortunately, the fact that $\kappa$ is constant (and equal to $\kappa_0$)
over a broad frequency band will allow the aforementioned cancellation of
the back-action, resulting in a QND measurement of speed.  To see
this, suppose that, instead
of measuring the output phase quadrature $\tilde{q}_2$, 
we use homodyne detection to measure a generic, 
frequency-independent quadrature 
of the output:
\begin{equation}
\tilde{q}_\Phi = \Delta \tilde{p}_1 \cos \Phi 
	+ (\Delta \tilde{p}_2 + \sqrt{2\kappa} \frac{h}{h_{SQL}} e^{i \psi} )
	\sin \Phi \,,
\end{equation}
where $\Phi$ is a fixed homodyne angle. 
Then from Eqs.~(\ref{qsout}), we infer that the noise in the signal,
expressed in GW strain units $h$, is 
\begin{equation}
h_n = \frac{h_{SQL}}{\sqrt{2\kappa}} e^{i\psi} 
[\tilde{p}_1 (\cot \Phi - \kappa) + \tilde{p}_2 ] \,.
\label{hn}
\end{equation}
By making
$\cot\Phi=\kappa_0 \equiv ({\rm constant\ value\ of\ \kappa\ at\ \omega 
\lesssim \Omega})$, the radiation pressure noise in $h_n$ will be 
cancelled in the broad band where $\kappa=\kappa_0$, thereby making
this a QND interferometer. 

We assume for now that ordinary vacuum enters the output port of the interferometer;
i.e., $\tilde{p}_1$ and $\tilde{p}_2$ are quadrature amplitudes for
ordinary vacuum (we will inject squeezed vacuum in Sec.~\ref{sec:sqvac}).  
This means [Eq.~(26) of KLMTV] that their (single-sided) spectral densities
are unity and their cross-correlations are zero, which, when combined
with Eq.~(\ref{hn}), implies a spectral
density of 
\beq
S_{h_n} 
    = (h_{SQL})^2 \xi^2 \,.
\label{simplespec}
\eeq
Here
\beq
\label{xidef}
\xi^2 \equiv \frac{(\cot\Phi-\kappa)^2+1}{2
    \kappa}
\eeq
is the fractional amount by which the SQL is beaten
(in units of squared amplitude).
This expression for $\xi^2$ is the same as that for the speed meters
in 
%Eq.~(\ref{xisq})
Paper~I [Eq.~(35)] 
and BGKT [Eq.~(40)], indicating the 
theoretical equivalency of these
designs.  In those papers, an optimization is given for the interferometer.
Instead of just using the results of that optimization, we shall carry out 
a more comprehensive study of it\footnote{%Eqs.~(\ref{qsout}) and (\ref{xwba}) 
It should be noted that the expressions given in
Sec.~\ref{sec:newlosslessmath} are accurate to 6\% or better 
over the frequency range
of interest.  To achieve 1\% accuracy, we expand to the next-highest
order.  The result can be expressed as a re-definition of
the sloshing frequency
$$
\Omega^2 \rightarrow \Omega'^2= \Omega^2 - \delta \delta_{\rm s}/2 \,,
$$
where 
$\delta_{\rm s} = c T_{\rm s}/2L$.  Then $\kappa$ retains the same functional
form:
$$
\kappa \rightarrow \kappa' = \frac{16 \omega_0 \delta W_{\rm circ}}{m c L 
((\Omega'^2-\omega^2)+\omega^2 \delta^2)} \,. 
$$
As a result, the optimization described in Sec.~\ref{sec:optimization} 
applies equally well to $\kappa'$ and $\Omega'$
as to the original $\kappa$ and $\Omega$.
\label{note:output}}.

\subsection{Optimization}
\label{sec:optimization}

\begin{figure}
\epsfig{file=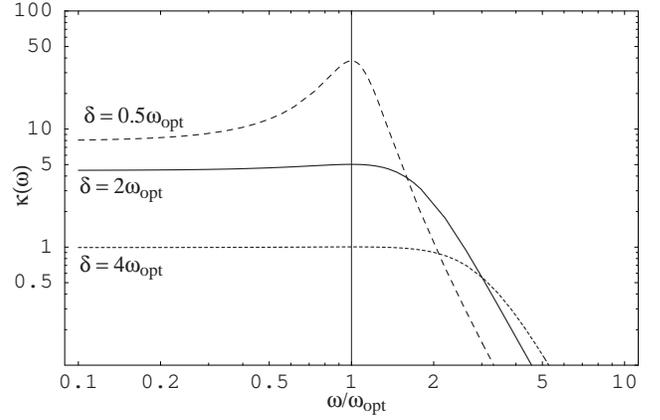,width=0.475\textwidth} 
\caption[Typical curves for the coupling constant $\kappa(\omega)$ in terms
of the optimal frequency $\omega_{\rm opt}$]
{The coupling constant $\kappa(\omega)$ with $\omega$ 
measured in units of $\omega_{\rm opt}$.  The solid curve
is determined by setting $\delta=2\omega_{\rm opt}$ and $\kappa_{\rm max}=5$ 
(this value of $\kappa_{\rm max}$ comes from specifying 
that we want to beat the SQL by a factor of 10; see Fig.~\ref{fig:differentdelta}).
If, in addition, we set $\omega_{\rm opt}=2 \pi \times 100~\rm Hz$, then 
all the parameters have been specified (due to the various relationships
between them) and are equal to the values
given in Table~\ref{table:newparams}.  If we maintain the same power but change 
$\delta$, then the only parameter of Table~\ref{table:newparams} that
is affected is $T_{\rm o}$. Examples of such a change are
shown for $\delta=0.5\omega_{\rm opt}$
and $\delta=4\omega_{\rm opt}$.
Note that these two choices of $\delta$ are more extreme than would
be desirable in practice, but they are shown here to illustrate more clearly
the effect on $\kappa$ of changing the ratio between $\delta$ and 
$\omega_{\rm opt}$. }
\label{fig:kappaomegaopt}
\end{figure}

The possible choices of
speed meter parameters can be investigated intuitively
by examining the behavior of
$\kappa$.  To aid us in our exploration, we choose 
(as in BGKT and 
%Chapter 2)
Paper~I) 
to 
express $|{\cal L}(\omega)|^2$
[Eq.\ (\ref{fancyL})] as
\begin{equation}
|{\cal L}(\omega)|^2 = (\omega^2-\omega_{\rm opt}^2)^2
	+ \delta^2 (\omega^2_{\rm opt} +\delta^2 /4) \,,
\end{equation}
where
\begin{equation}
\omega_{\rm opt} = \sqrt{\Omega^2 - \delta^2/2} \,,
\end{equation}
is the interferometer's ``optimal frequency," i.e., the frequency
at which $|{\cal L}(\omega)|$ reaches its minimum.  Combining 
with Eq.~(\ref{kappa}), we obtain
\beq
\label{kappasimple}
\kappa=\frac{\Omega_{\rm I}^3\,\delta}{(\omega^2-\omega_{\rm
opt}^2)^2+\delta^2 (\omega^2_{\rm opt} +\delta^2 /4)}\,,
\eeq
where 
\beq
\Omega_{\rm I}^3 \equiv \frac{16\omega_0 W_{\rm circ}}{mLc}
\label{kappafreqscale}
\eeq
is a frequency scale related to the circulating power. 
At $\omega_{\rm opt}$, $\kappa$ reaches its
maximum (see Fig.~\ref{fig:kappaomegaopt})
\beq
\label{kappamax}
\kappa_{\rm max}
=\frac{\Omega_{\rm I}^3}{\delta (\omega_{\rm opt}^2+\delta^2/4)}\,.
\eeq
By setting
\beq
\label{phifixed}
\cot\Phi=\kappa_{\rm max}\,,
\eeq
we get the maximum amount by which a speed meter can beat the SQL
\beq
\label{ximin}
\xi_{\rm min}^2=\frac{1}{2\kappa_{\rm max}}=\frac{\delta(\omega_{\rm
opt}^2+\delta^2/4)}{2\Omega_{\rm I}^3}\,.
\eeq

\begin{figure}[t]
\epsfig{file=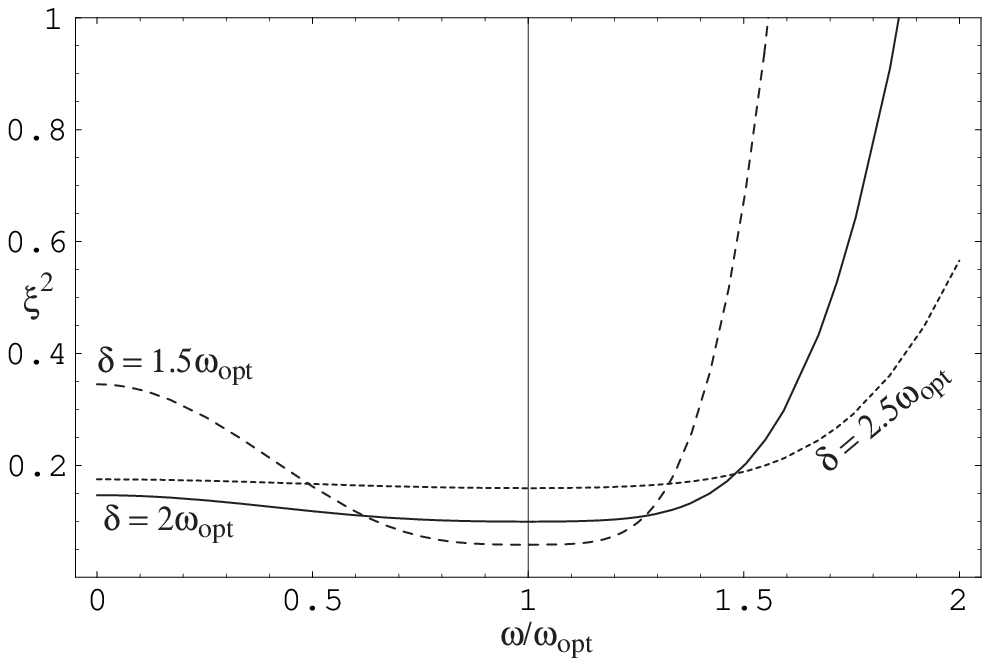,width=0.475\textwidth} 
\caption[Squared amount by which the speed meter beats the SQL]
{The squared amount by which the speed meter beats the SQL with
a given circulating power, which is determined by setting
(for the solid curve) $\xi^2_{\rm min}=0.1$ and the condition (\ref{doublexi}).
%$\delta= 2\omega_{\rm opt}$.
Note that the requirement on $\xi_{\rm min}^2$ sets the power relative to
the SQL power $W_{\rm circ}^{SQL}$, the value of which is dependent 
on $\omega_{\rm opt}$. (For $\omega_{\rm opt}=100~\rm Hz$, we have 
$W_{\rm circ} = 8~\rm MW$.) If we hold the power fixed and change $\delta$
to $1.5\omega_{\rm opt}$ and $2.5\omega_{\rm opt}$, we get the other 
two curves.}
\label{fig:differentdelta}
\vspace{.2cm}
%\end{figure}
%\begin{figure}
\epsfig{file=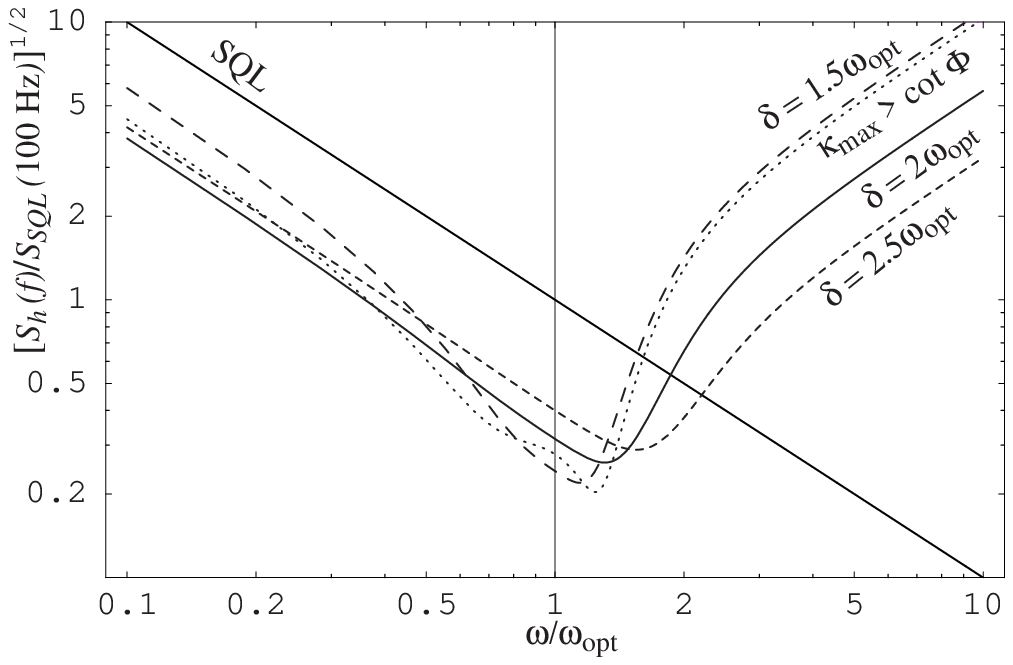,width=0.475\textwidth} 
\caption[Noise curves corresponding to the $\xi^2$ curves in preceding plot]
{Noise curves corresponding to the $\xi^2$ curves in
Fig.~\ref{fig:differentdelta}, the caption of which describes
the parameters used here as well.  
The dotted line is an example of a noise curve for which $\kappa$
is not quite flat and $\cot \Phi$ was chosen to be slightly smaller 
than $\kappa_{\rm max}$ 
(see the end of Sec.~\ref{sec:optimization} for details.)}
\label{fig:differentdeltanoise}
\end{figure}

As $\omega$ differs from $\omega_{\rm opt}$ in either direction,
$\kappa$ decreases from $\kappa_{\rm max}$. This causes the noise to
increase since (i) the term $(\cot\Phi-\kappa)^2$ in the numerator 
of $\xi^2$ [Eq.~(\ref{xidef})] 
increases and (ii) the denominator of $\xi^2$ decreases. 
In order to have broadband performance, we should make the 
peak of $\kappa(\omega)$ as flat as
possible.  As we can see from both Eq.~(\ref{kappasimple}) and
Fig.~\ref{fig:kappaomegaopt}, the shape of the peak can be adjusted by
changing $\delta$: for the same optical power, a larger $\delta$ means
a wider peak but a smaller maximum. 
Therefore, changing $\delta$ is one method of balancing 
sensitivity against bandwidth. Some examples are shown in
Figs.~\ref{fig:kappaomegaopt}, \ref{fig:differentdelta}, and \ref{fig:differentdeltanoise}, where 
$\kappa(\omega)$, $\xi^2(\omega)$, and $S_h(\omega)$, respectively, 
are plotted for configurations with
the same $\omega_{\rm opt}$ and optical power $W_{\rm circ}$, 
but with several values of $\delta$. 

To be more quantitative, a simple analytic form for $\xi^2(\omega)$ can be
obtained by inserting Eqs.~(\ref{kappasimple}), (\ref{kappamax}), and 
(\ref{ximin}) into Eq.~(\ref{xidef}) to get
\beq
\xi^2(\omega)
=
\left[1+\Delta+\frac{1}{4\xi_{\rm
min}^4}\frac{\Delta^2}{(1+\Delta)}\right]\xi_{\rm min}^2\,.
\eeq
Here 
\beq
\label{Deltadef}
\Delta\equiv\frac{\left(\omega^2-\omega_{\rm
opt}^2\right)^2}{\delta^2(\omega_{\rm opt}^2+\delta^2/4)}\,
\eeq
is a dimensionless offset from the optimal frequency $\omega_{\rm opt}$. 
From Eq.~(\ref{Deltadef}), it is evident that $\Delta$, and thus
$\xi^2$, are the same for $\omega=0$ and $\omega=\sqrt2
\omega_{\rm opt}$ [see also Eq.~(47) of BGKT or 
%Eq.~(\ref{range})].
Eq.~(49) of Paper~I].
For definiteness, let us impose that
\beq
\label{doublexi}
\xi^2(0)=\xi^2(\sqrt2\omega_{\rm opt})=\frac{3}{2}\xi^2_{\rm min}
%\xi^2(0)=\xi^2(\sqrt2\omega_{\rm opt})=2\xi^2_{\rm min}\,,
\eeq
as is done by BGKT.
For $\xi_{\rm min}^2=0.1$, this gives $\delta=1.977\omega_{\rm
opt}\approx 2\omega_{\rm opt}$ (as assumed in BGKT and 
%Chapter 2).
Paper~I).  
Plugging these numbers into Eq.~(\ref{ximin}) 
and combining with Eq.~(\ref{kappafreqscale}) gives
%In this case, Eq.~(\ref{wcirc}) requires 
\bea
W_{\rm circ} && ({\delta=2\omega_{\rm opt}})
=\frac{mLc\,\omega_{\rm opt}^3}{8\omega_0\xi_{\rm min}^2} \nonumber \\
&&\simeq 8.4\,{\rm MW}
\left(\frac{\omega_{\rm opt}}{2\,\pi\times 100\,{\rm Hz}} \right)^3
\left(\frac{m}{40\,{\rm kg}} \right) 
\nonumber \\
&& \quad \times 
\left(\frac{L}{4000\,{\rm km}} \right) 
\left(\frac{1.78\times10^{15}\,{\rm Hz}}{\omega_0}\right) 
\left( \frac{0.1}{\xi_{\rm min}^2} \right) \,.
\label{wcircex}
\eea
Therefore, when $\omega_{\rm opt}$ is chosen at $2\pi\times100\,{\rm
Hz}$, this speed meter (with $\delta=2\omega_{\rm opt}$) requires $W_{\rm
circ} \simeq 8.4\,{\rm MW}$ to beat the SQL by a factor of 10
in power ($\xi_{\rm min}^2=0.1$). 
%{\it However,} significantly 
%changing $\xi^2_{\rm min}$ in the above equation (without changing
%the ratio between $\delta$ and $\omega_{\rm opt}$) will change the 
%wide-band performance of the interferometer, since there is some
%``hidden" power-dependence in Eq.~(\ref{doublexi}). 
[Note that, keeping $\delta=2\omega_{\rm opt}$, the speed meter reaches
the SQL with $W_{\rm circ}^{\rm SQL}=840\,{\rm kW}$, comparable to
the value given by KLMTV Eq.~(132) for conventional 
interferometers with 40-kilogram 
test masses.] 
The $\xi^2$ and $S_h$ curves
for this configuration are plotted as solid lines
in Fig.~\ref{fig:differentdelta} and
\ref{fig:differentdeltanoise}, respectively.  

Please note that Eq.~(\ref{wcircex}) should be applied with caution
because significantly 
changing $\xi^2_{\rm min}$ in the above equation (without changing
the ratio between $\delta$ and $\omega_{\rm opt}$) will change the 
wide-band performance of the interferometer, since there is some
``hidden" power dependence in Eq.~(\ref{doublexi}). 
To determine the behavior of the speed meter with significantly higher 
power or lower
$\xi_{\rm min}^2$ {\it while maintaining the same wideband performance}, 
we must re-apply the requirement (\ref{doublexi})
to determine the appropriate ratio between $\delta$ and $\omega_{\rm opt}$.  
For example, solving
Eqs.~(\ref{ximin}) and (\ref{doublexi}) simultaneously for $\xi^2_{\rm min}$
and $\delta$, with chosen values $W_{\rm circ}=20~{\rm MW}$ and 
$\omega_{\rm opt}=2 \pi \times 100~{\rm Hz}$, gives
$\delta=2.334 \omega_{\rm opt}$ and $\xi_{\rm min}^{-2} \simeq 17$. 
Keeping this in mind, a general expression for the circulating power is 
\bea
\label{wcirc}
W_{\rm circ}
&=&\frac{mLc(\omega_{\rm opt}^2+\delta^2/4)\delta}{32\omega_0\,\xi_{\rm min}^2} 
\nonumber \\
&=&\frac{209\,{\rm kW}}{\xi_{\rm min}^2}
	\left[\frac{(\omega_{\rm opt}^2+\delta^2/4)\delta}
				{(2\pi\times100\,{\rm Hz})^3}\right]
\nonumber \\
&&\times
\left(\frac{m}{40\,{\rm kg}} \right)
\left(\frac{L}{4000\,{\rm km}} \right) 
\left(\frac{1.78\times10^{15}\,{\rm Hz}}{\omega_0}\right)\,,
\eea
where the relationship between $\delta$ and $\omega_{\rm opt}$ determines
whether the noise curve is deep but narrow or wide but shallow [with
the requirement (\ref{doublexi}) giving the latter].

\begin{figure}
\epsfig{file=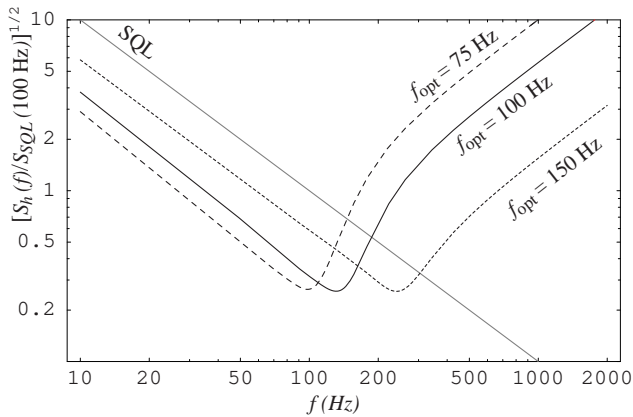,width=0.475\textwidth} 
\caption[Noise curves for varying optimal frequencies]
{Noise curves for varying optimal frequencies.  The
solid curve has $f_{\rm opt}=100$ Hz and is identical
to the solid curve of Fig.~\ref{fig:differentdeltanoise}.
Maintaining the same power and the condition imposed by 
Eq.~(\ref{doublexi}),
we show two examples of noise curves with other optimal
frequencies, specifically 
$f_{\rm opt}= 75$ Hz and 
$f_{\rm opt}= 150$ Hz. }
\label{fig:omegaopt}
\end{figure}

So far, we have only changed $\delta$ to modify the performance of the
speed meter.  Another method is to change $\omega_{\rm opt}$.  In this case, 
the shape of the noise curve changes very little, but the minima occur 
at different frequencies, causing the interferometer to have either broader
bandwidth or higher sensitivity (relative to the SQL).  This is shown in
Fig.~\ref{fig:omegaopt}.
Maintaining condition (\ref{doublexi}) 
with $\omega_{\rm opt}$ chosen at
$2\pi\times 150$\,Hz, we get a broader but
shallower curve (short dashes); this configuration 
beats the SQL by a factor of $\xi_{\rm min}^{-2} \sim 4.7$, up to
$f \sim 240\,{\rm Hz}$.  With $\omega_{\rm opt} = 2\pi\times 75$\,Hz, we get 
a narrower but deeper curve (long dashes), which
beats the SQL by a factor of 
$\xi_{\rm min}^{-2} \sim 17$, up to $f \sim 100\,{\rm Hz}$.
The power was kept fixed at $W_{\rm circ}=8.2\,{\rm MW}$.

One more potential optimization method is to choose a
$\kappa$ with a peak that is not quite flat and then choose a $\cot\Phi$
that is slightly smaller than $\kappa_{\rm max}$.  This will give a 
wider bandwidth on either side of $\omega_{\rm opt}$, at the price of
decreased sensitivity at the region near $\omega_{\rm opt}$ (see
dotted line in Fig.~\ref{fig:differentdeltanoise}).

For simplicity, we will choose a typical (but somewhat arbitrary) 
set of parameters for the lossless 
interferometer of Fig.~\ref{fig:oneport}.  
These values, given in Table \ref{table:newparams},
will be used (except as otherwise noted) for subsequent plots and
calculations comparing this speed-meter design to other configurations.

\begin{table}[t]
\caption[Three-arm speed-meter interferometer parameters and their fiducial values]
{Three-arm speed-meter interferometer parameters and their fiducial values,
as used throughout except where other parameters are specified.}
%\begin{minipage}{6in}
\begin{tabular}{lllll}
%\hline
%\hline
Parameter&Symbol&Fiducial Value\\
\hline
carrier frequency & $\omega_0$ & $1.78 \times 10^{15} {\rm s}^{-1}$ \\
mirror mass & $m$ & 40 kg \\
arm length & $L$ & 4 km \\
sloshing mirror transmissivity & $T_{\rm s}$ & 0.0008 \\
output mirror transmissivity & $T_{\rm o}$ & 0.017 \\
end mirror transmissivity & $T_{\rm e}$ & $2 \times 10^{-5}$ \\
%power-recycling mirror transmissivity && $T_{\rm p}$ & $10^{-5}$ \\
internal and RSE mirror trans. & $T_{\rm i}$ & 0.005 \\
optimal frequency & $\omega_{\rm opt}$ & $2\pi \times 100~\rm Hz$ \\
\hline
sloshing frequency & $\Omega$ & $2\pi \times 170~\rm Hz$\\
extraction rate (half-bandwidth) & $\delta$ & $2\pi \times 200~\rm Hz$ \\
SQL circulating power & $W_{\rm circ}^{SQL}$ & 820 kW \\
\end{tabular}
%\end{minipage}
%\begin{tabular}{c}
%\hline
%\hline
%\hspace*{5.8in}\\
%\end{tabular}
\label{table:newparams}
%\vspace*{-.2in}
\end{table}

\subsection{Discussion of Three-Cavity Speed-Meter Design}
\label{sec:basicdisc}

In this section, we discuss how the three-cavity speed-meter design compares to the 
two-cavity design presented in 
%Chapter 2,
Paper~I, 
focusing on the three major problems of 
that design: it required (i) a high circulating power,
(ii) a large amount of power coming out of the interferometer
with the signal, and (iii) an exorbitantly high input laser power.  

With the three-cavity speed meter, we are able to replicate the
performance of the two-cavity design in 
%Chapter 2,
Paper~I, 
but without the exorbitantly high
input power. 
The reason why our three-cavity speed meter 
does not need a high input power is the same as for 
conventional interferometers: in both cases, the 
excited cavities are fed directly by the laser.  According to Bose statistics,
carrier photons will be ``sucked" into the cavities, 
producing a strong amplification.
This was not the case in the two-cavity speed meter of 
%Chapter 2.
Paper~I.  
There, an essentially
empty cavity stood between the input and the excited cavity, thereby 
thwarting Bose statistics and resulting in
a required input laser power much greater than the power that 
was circulating in the 
excited cavity (see 
%Chapter 2
Paper~I
for more details).  
In this paper, 
%In this chapter, 
we have
returned to a case where the laser is driving an excited cavity directly,
thereby allowing the input laser power to be small relative to the circulating
power.

Because the cavity from which we are reading out the signal
does not contain large amounts of carrier light (by contrast with the
two-cavity design), this three-cavity speed meter does
not have large amounts of power exiting the interferometer with the
velocity signal, unlike the two-cavity design.  By making use of the 
different modes of the Michelson
interferometer, we have solved the problem of the exorbitantly high input power
and the problem of the amount of light that comes out of the interferometer.

The problem of the high circulating power $W_{\rm circ}$, 
unfortunately, is not solved by the three-cavity design.  This is 
actually a common characteristic of ``external-readout" interferometer 
designs capable of beating the SQL.  The reason for this high power
is the energetic quantum limit (EQL), which was first derived for
gravitational-wave interferometers by 
Braginsky, Gorodetsky, Khalili and Thorne
\cite{highpower}.  The EQL 
arises from the phase-energy uncertainty principle
\beq
\Delta E \Delta \phi \ge \frac{\hbar \omega_0}{2}\,,
\eeq 
where $E$ is the stored energy in the interferometer and $\phi$ is
the phase of the light. 
The uncertainty $\Delta E$ of the stored light energy 
during the measurement process must
be large enough to allow a small uncertainty $\Delta \phi$ in the 
stored light's optical
phase, in which the GW signal is contained. For an interferometer with
coherent light (so $\Delta E= \hbar \omega_0 \sqrt{E/\hbar \omega_0}$), 
the EQL dictates that the energy stored in
the arms must be larger than 
\beq
\label{EQL}
E_{\rm \xi}\sim\frac{m L^2\omega^2\Delta \omega}{4\omega_0\xi^2}\,
\eeq
in order to  beat the SQL by a factor of $\xi$
near frequency  $\omega$ with a bandwidth $\Delta \omega$ 
(Eq.~(1) of Ref.~\cite{opticallever} and 
Eq.~(29) of Ref.~\cite{highpower}).  
In a
broadband configuration with $\Delta \omega \sim \omega$, we have  
\beq
\label{EQLbb}
E_{\rm \xi}\sim\frac{m L^2\omega^3}{4\omega_0\xi^2}\,.
\eeq 
For comparison, in the broadband regime of the speed meter, we have,
from Eq.~(\ref{ximin}), 
\beq
\label{SMbb}
\xi_{\rm min}^2=\frac{m L^2 \delta(\omega_{\rm opt}^2+\delta^2/4)}{4\,E
\omega_0}
\sim 
\frac{m L^2 \omega_{\rm opt}^3}{4\,E\omega_0}\,,
\eeq
where the stored energy is $E=2 W_{\rm circ} L/c$. Comparison
between Eqs.~(\ref{EQLbb}) and (\ref{SMbb}) confirms
that our speed meter is EQL limited. 

As a consequence of the EQL, designs with coherent light will all
require a similarly high circulating power in order to achieve a
similar sensitivity. Moreover, given the sharp dependence $E\propto \omega^3$,
this circulating power problem will become much more severe when one wants
to improve sensitivities at high frequencies. 

Nevertheless, the EQL in the form
(\ref{EQL}) above only applies to coherent light. 
Using nonclassical light will enable the interferometer to circumvent
it substantially.  One possible method was invented by Braginsky, Gorodetsky,
and Khalili \cite{symphotonicstates} using a special optical topology and
intracavity signal extraction.  A more conventional solution for our
external-readout interferometer is to inject squeezed light into the 
dark port, as we shall discuss in Sec.~\ref{sec:sqvac} (and as was
also discussed in the original paper \cite{highpower} on the EQL).

\section{Squeezed Vacuum and FD Homodyne Dectection}
\label{sec:modifications}

In this section, we discuss two modifications to the three-cavity speed-meter
design analyzed in Sec.~\ref{sec:newlosslessmath}.  
This discussion applies to both the simple and practical
versions, shown in Figs.~\ref{fig:oneport} and \ref{fig:oneport3IM};
the modifications are shown in Fig.~\ref{fig:threearmfull}.
The first modification is to inject 
squeezed vacuum (with fixed squeeze angle) into 
the output port of the speed meter, as shown in 
Fig.~\ref{fig:threearmfull}.  This will reduce the amount of power 
circulating in the interferometer.  The second modification, also shown in Fig.~\ref{fig:threearmfull}, is the introduction of two
filter cavities on the output, which allow us to perform 
{\it frequency-dependent} homodyne detection (described in KLMTV) 
that will dramatically
improve the performance of the speed meter at frequencies $f \gtrsim f_{\rm opt}$.

\subsection{Injection of Squeezed Vacuum into Dark Port}
\label{sec:sqvac}

Because the amount of circulating power required by our speed meter
remains uncomfortably large, it is desirable to reduce it by injecting 
squeezed vacuum into the dark port. The idea of
using squeezed light in gravitational-wave interferometers 
was first conceived by Caves \cite{CavesSqVac}
and further developed by Unruh
\cite{Unruhsqueezedvacuum} and KLMTV. 
We shall start in this section with a
straightforward scheme that will decrease the effective circulating
power without otherwise changing the speed meter performance.

As discussed in Sec.~IV B and Appendix A of KLMTV, a squeezed input state is
related to the vacuum input state (assumed in Sec.~\ref{sec:newlosslessmath}) 
by a unitary squeeze operator $S(R,\lambda)$
[see Eqs.~(41) and (A5) of KLMTV]
\beq
|{\rm in}\rangle = S(R,\lambda)|0\rangle\,.
\eeq
Here $R$ is the squeeze amplitude and $\lambda$ is the squeeze
angle, both of which in principle can depend on sideband frequency. However, the
squeezed light generated using nonlinear crystals \cite{XWK87,GSYL87} has
frequency-independent $R$ and $\lambda$ in our frequency band of interest, 
i.e., $f < 10$\,kHz \cite{Kip}; and in this section, we shall assume 
frequency independence.

The effect of input squeezing is most easily understood 
in terms of the following unitary transformation,
\begin{mathletters}
\bea
|{\rm in}\rangle &\rightarrow& S^{\dagger}(R,\lambda) |{\rm in}\rangle
= |0\rangle\, \\
\tilde{p}_j&\rightarrow& S^{\dagger}(R,\lambda)
\tilde{p}_j S(R,\lambda)\,, \\
\tilde{q}_j&\rightarrow& S^{\dagger}(R,\lambda)
\tilde{q}_j S(R,\lambda)\,,
\eea
\end{mathletters}
where $j=1,2$.  This 
brings the input state back to vacuum and transforms the input
quadratures into linear
combinations of themselves, in a
rotate-squeeze-rotate way [Eq.~(A8) of KLMTV, in matrix form]:
\bea
\label{squeezetransf}
\left( \begin{array}{c} \tilde{p}_1 \\ 
	\tilde{p}_2 \end{array} \right) &\rightarrow & 
\left(
\begin{array}{c}
\tilde{p}_{1s} \\
\tilde{p}_{2s}
\end{array}
\right) =
 S^\dagger(R,\lambda)
\left(
\begin{array}{c}
\tilde{p}_1 \\
\tilde{p}_2
\end{array}
\right)
S(R,\lambda) \nonumber \\
&& \quad =
\left(
\begin{array}{rr}
\cos\lambda & -\sin\lambda \\
\sin\lambda & \cos\lambda
\end{array}
\right) 
\left(
\begin{array}{cc}
e^{-R} & 0 \\
    0  & e^{R}
\ea\right)
\nonumber \\
&&\quad \quad \quad \times 
\left(\ba{rr}
\cos\lambda & \sin\lambda \\
-\sin\lambda & \cos\lambda
\end{array}
\right)
\left(
\begin{array}{c}
\tilde{p}_1 \\
\tilde{p}_2
\end{array}
\right)\,.
\eea
In particular, the GW noise can be calculated by using the squeezed noise operator [Eq.~(29) of KLMTV] 
\begin{equation}
h_{ns} = S^\dagger (R,\lambda) h_n S(R,\lambda) \,,
\label{squeezedef}
\end{equation}
and the vacuum state.

A special case---the case that we want---occurs when $R={\rm constant}$ 
and $\lambda=\pi/2$.  Then there is no
rotation between the quadratures but only a 
frequency-independent squeezing or stretching,
\begin{mathletters}
\bea
\tilde{p}_1 &\rightarrow& \tilde{p}_{1s} = e^{R} \tilde{p}_1 \,,\\
\tilde{p}_2 &\rightarrow& \tilde{p}_{2s} = e^{-R} \tilde{p}_2 \,.
\eea
\end{mathletters}
Consequently, Eqs.~(\ref{qsoutKimbleform}) for the output quadratures
$\tilde{q}_{1,2s} = S^\dagger (R,\pi/2) \tilde{q}_{1,2}
S(R,\pi/2) $ are transformed into
\begin{mathletters}
\bea
\tilde{q}_{1s}&=& e^R \tilde{p}_1 e^{2i\psi} \\
\tilde{q}_{2s}&=&  e^{-R} \biggl[ \left( \tilde{p}_2 
		- \kappa e^{2 R} \tilde{p}_1\right) e^{2i\psi}
	+ \sqrt{\kappa e^{2 R}}\frac{\tilde{h}}{h_{SQL}}e^{i\psi}\biggr]\,.
\eea
\end{mathletters}
The corresponding noise can be put into the same form as Eq.~(\ref{hn}),
\beq
h_{ns}
	=\frac{h_{SQL}}{\sqrt{\kappa_{\rm eff}}} e^{i\psi}
		 \left[\tilde{p}_1(\cot\Phi_{\rm eff}-\kappa_{\rm eff})
		 	+\tilde{p}_2\right]\,,
\eeq
with
\beq
\cot\Phi_{\rm eff}\equiv e^{2 R} \cot\Phi\,, \quad
\kappa_{\rm eff}\equiv e^{2 R} \kappa \,.
\label{sqphi}
\eeq
Since $\kappa$ is proportional to the circulating
power [see Eqs.~(\ref{kappa})], gaining a factor $e^{2R}$ in
$\kappa$ is equivalent to gaining this factor in $W_{\rm circ}$.

In other words, by injecting squeezed vacuum with squeeze factor
$e^{2R}$ and squeeze angle $\lambda=\pi/2$ into the interferometer's
dark port, we can achieve precisely the same interferometer performance
as in Sec.~\ref{sec:newlosslessmath}, but with a circulating light power
that is lower by $W_{\rm circ, SISM} = e^{-2R} W_{\rm circ, OSM}$.  (Here
``SISM" means ``squeezed-input speed meter" and ``OSM" means ``ordinary
speed meter."  Since squeeze factors $e^{-2R}\sim 0.1$ are likely to be
available in the time frame of LIGO-III \cite{Kimble}, 
this squeezed-input speed meter
can function with $W_{\rm circ, SISM} \simeq 0.1 W_{\rm circ, OSM}$.

\subsection{Frequency-Dependent Homodyne Detection}
\label{sec:sqvacFDdetect}

One can take further advantage of squeezed light by using
frequency-dependent (FD) homodyne
detection at the interferometer output \cite{vo1,vo2,vo3,vo4,vo5}.
As KLMTV have shown, FD homodyne detection can be achieved by sending
the output light through one or more optical filters 
(as in Fig.~\ref{fig:threearmfull}) and then performing ordinary
homodyne detection.
If its implemention is 
feasible, FD homodyne detection will dramatically improve the speed meter's
sensitivity at high frequencies (above $f_{\rm opt}=100$~Hz).
Note that the KLMTV design that used FD homodyne detection was called a ``variational-output" interferometer; consequently, we shall use the
term ``variational-output speed meter" to refer to our speed meter 
with FD homodyne detection.  Continuing the analogy, when we have 
both squeezed-input and FD homodyne detection, we will use the term ``squeezed-variational speed meter." 
%but it should be noted that the
%squeezed-variational position meter of KLMTV uses frequency-dependent 
%squeezed vacuum, but our squeezed-variational speed meter only needs
%frequency-independent squeezed vacuum.
The following
discussion is analogous to Secs.~IV and V of KLMTV.

For a generic frequency-dependent\footnote{For generality 
of the equations, we allow the squeeze angle and the 
the homodyne phase both to be frequency dependent, but
the squeeze angle will be fixed (frequency independent) 
later in the argument 
[specifically, in Eq.~(\ref{fdhomodyne})].} squeeze 
angle $\lambda (\omega)$ and homodyne
detection phase $\Phi (\omega)$, we have, for the squeezed noise operator
[Eqs.~(\ref{squeezedef}) and (\ref{squeezetransf})],
\begin{eqnarray}
&& h_{ns} =  - \frac{h_{SQL}}{\sqrt{\kappa}} e^{i \psi }
	\sqrt{1 + \tilde{\kappa}^2} 
	\nonumber \\
	&& \;\; \times 
	\biggl(\tilde{p}_1  \bigl\{ \cosh R \cos \tilde{\Psi} 
		- \sinh R \cos \bigl[ \tilde{\Psi} 
			- 2 ( \tilde{\Psi} + \lambda ) \bigr] \bigr\} \nonumber \\
	&& \;\;
	- \tilde{p}_2 \bigl\{ \cosh R \sin \tilde{\Psi} 
		- \sinh R \sin \bigl[ \tilde{\Psi} - 2 ( \tilde{\Psi} + \lambda ) \bigr] \bigr\} \biggr) \,,
\label{hns}
\end{eqnarray}
where 
\begin{equation}
\cot \tilde{\Psi} \equiv \tilde{\kappa} \equiv \kappa - \cot \Phi \,.
\end{equation}
The corresponding noise spectral density [computed by using the ordinary
vacuum spectral densities, $S_{\tilde{p}_1}=S_{\tilde{p}_2}=1$ and
$S_{\tilde{p}_1 \tilde{p}_2}=0$, in Eq.~(\ref{hns})] is
\bea
\label{shns}
S_{h} 
	&=& \frac{(h_{SQL})^2}{\kappa} (1 + \tilde{\kappa}^2) 
	\nonumber \\
	  && \qquad \times
	  \biggl\{ e^{-2R}  + \sinh 2R [1 - \cos 2(\tilde{\Psi}+\lambda)] 
	\biggr\} \,.
\eea
Note that these expressions are analogous to KLMTV Eqs.~(69)--(71) for a 
squeezed-variational interferometer (but the frequency dependence of their
$\cal K$ is different from that for our $\kappa$). From
Eq.~(\ref{shns}), $S_h$ can be no smaller than the case when 
\beq
\label{kphi}
\tilde{\kappa}=0\,,\quad \cos 2(\tilde{\Psi}+\lambda)=1\,.
\eeq
The optimization conditions (\ref{kphi}) are satisfied when 
\beq
\label{fdhomodyne}
\cot\Phi=\kappa\,,\quad \lambda=\pi/2\,,
\eeq
which corresponds to frequency-dependent homodyne detection on the
(frequency-independent) squeezed-input speed meter discussed in the
previous section. 

As it turns out, the condition $\cot \Phi = \kappa$
can readily be achieved by the family of
two-cavity optical filters invented by KLMTV and discussed in their
Sec.~V and Appendix~C. We summarize and generalize their main results
in our Appendix~\ref{app1}. 
The two filter cavities are both Fabry-Perot
cavities with (ideally) only one transmitting mirror. They are
characterized by their bandwidths, $\delta_{J}$, (where $J={\rm I,\,II}$ denote
the two cavities) and by their
resonant frequencies,  $\omega_0+\xi_{J} \delta_{J}$ (the ones nearest
$\omega_0$). The output light from the squeezed-input speed meter
is sent through the two filters, and then a homodyne detection
with frequency-independent phase $\rm \theta$ is performed on it.  
%(Faraday
%isolators are inserted into the output--filter I--filter II optical
%train to avoid optical feedback.)

\begin{figure}
\epsfig{file=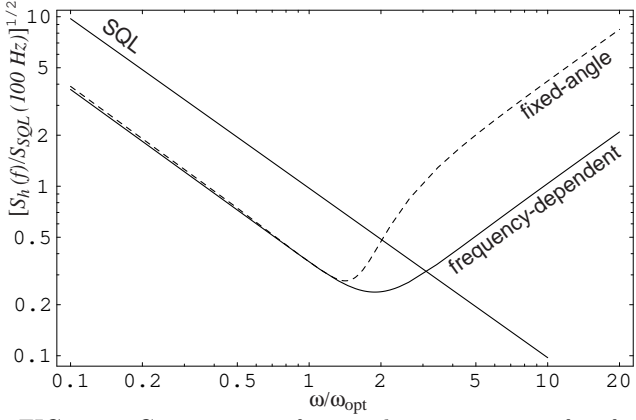,width=0.475\textwidth} 
\caption[Comparison of typical noise curves for 
frequency-dependent and fixed-angle homodyne detection]
{Comparison of typical noise curves for 
frequency-dependent and fixed-angle homodyne detection.  
The FD homodyne angle $\Phi (\omega)$ is that of Eqs.~(\ref{fdhomodyne})
and (\ref{kappaexample}); the fixed homodyne angle $\Phi$ is that of
Eq.~(\ref{sqphi}); the circulating power is $e^{-2R}$ times that of 
Table~\ref{table:newparams}; and 
all other parameters are identical for the two interferometers
and are given in Table~\ref{table:newparams}.}
\label{fig:fdnc}
\end{figure}

For the squeezed-variational speed meter
(shown in Fig.~\ref{fig:threearmfull}) 
with the parameters in Table~\ref{table:newparams}, plus 
$\xi_{\rm min}^2=0.1$, $\delta=2\omega_{\rm opt}$,
$\Lambda^4=4\,\omega_{\rm opt}^4$, and $e^{-2R}=0.1$, we
have 
\beq
\label{kappaexample}
\kappa
	=\frac{4\,\omega_{\rm opt}^4}{(\omega^2-\omega_{\rm opt}^2)^2
		+8\,\omega_{\rm opt}^4}\,
\eeq
and the required filter and detection configuration is
$\xi_{\rm I}=1.7355$, 
$\delta_{\rm I}=2\pi\times 91.57\,{\rm Hz}$, 
$\xi_{\rm II}=-1.1133$, 
$\delta_{\rm II}=2\pi\times 114.3\,{\rm Hz}$, and $\theta=\pi/2$. 
[These values are reached by solving Eqs.~(C4) of KLMTV, or by
using the simpler method described in Appendix~\ref{app1} of this
paper.]  The
resulting performance is plotted in Fig.~\ref{fig:fdnc}. 
Note the substantial improvement at $\omega \gtrsim \omega_{\rm opt}$.

In the case of position-meter interferometers with optical filters (the
interferometers analyzed by KLMTV), the
optical losses due to the filter cavities contribute significantly
to the noise spectral density and drastically reduce the ability
to beat the SQL. It turns out that the squeezed-variational speed meter is
less sensitive to such losses,
as we shall see in Sec.~\ref{sec:newlossy}.

\section{Optical Losses}
\label{sec:newlossy}

In order to understand the issue of optical losses in this speed meter,
we shall start by addressing its {\it internal} losses.
These include scattering and absorption at each optical element, 
finite transmissivities of the end mirrors, and imperfections
of the mode-matching between cavities.  The effect of {\it external} losses (i.e.,
losses in the detection system and any filter cavities) will be discussed 
separately.  Note that the analysis in this section includes the
internal and RSE mirrors, so it applies primarily to the speed meter
designs in Figs.~\ref{fig:oneport3IM} and \ref{fig:threearmfull}. 

\subsection {Internal losses}
\label{sec:internalloss}

In this subsection, we will consider only noise resulting from losses
associated with optical elements inside the interferometer.  These 
occur
\begin{itemize}
\item in the optical elements: arm cavities, sloshing cavity, extraction mirror,
port-closing mirror, beam splitter, RSE mirror;
%\begin{itemize}
%\item arm cavities,
%\item sloshing cavity,
%\item extraction mirror,
%\item port-closing mirror,
%\item beam splitter,
%\item RSE mirror;
%\end{itemize}
\item due to mode-mismatching\footnote{According to our simple analysis
in Appendix \ref{app:modemismatch}, this effect will be 
insignificant in comparison with
the losses in the optical elements, so we shall ignore it.}; and
\item due to the imperfect matching of the transmissivities of the 
RSE and internal mirrors\footnote{This effect is negligibly small
so we shall ignore it; see Appendix~\ref{sec:mirrormismatch} for details.}.
\end{itemize}
Since the optical losses will dominate, we focus only on that type of 
loss here. 
The loss at each optical element will decrease the amplitude of the sideband
light (which carries the gravitational-wave information) and will 
simultaneously
introduce additional vacuum fluctuations into the optical
train.  Schematically, for some sideband $\tilde{a}(\omega)$, the loss
is described by
\begin{equation}
\label{includeloss}
\tilde{a}(\omega )\rightarrow \sqrt{1-{\cal E}(\omega )}\,\tilde{a}(\omega )
	+\sqrt{{\cal E}(\omega )}\,\tilde{n}(\omega )\, ,
\end{equation}
where \( \cal {E} \) is the (power) loss coefficient, and \(\tilde{n}(\omega ) \)
is the vacuum field entering the optical train at the loss point. 

It should be noted that there are various methods of grouping these
losses together in order to simplify calculations.  For example,
we combine all of the losses occurring in the arm (or sloshing) cavities, 
into one loss coefficient of ${\cal L} \sim 20 \times 10^{-6}$
[according to KLMTV Eq.~(93)].  Then we assume that the end
mirrors have transmissivity $T_{\rm e}=2\times 10^{-5}$, thereby
absorbing all of the arm losses into one term [see KLMTV Eq.~(B5) and preceding
discussion].

Assuming that the noise entering at the end mirrors of the arm cavities
is denoted $\tilde{n}_{e1,2}$ and $\tilde{n}_{n1,2}$ for the east and 
north arms respectively, at the end mirror of the 
sloshing cavity $\tilde{s}_{1,2}$, at the port-closing mirror
$\widetilde{w}_{1,2}$, and at the RSE mirror $\widetilde{m}_{n1,2}$
and $\widetilde{m}_{s1,2}$ [representing the losses described in the
previous paragraph; see Appendix~\ref{app:RSEcavity} for details], 
the output of the lossy three-cavity speed-meter 
system (Fig.~\ref{fig:oneport3IM}; the simplified and practical versions
are no longer equivalent, since there will be additional losses due to 
the presence of the internal and RSE mirrors)
is
\begin{mathletters}
\begin{eqnarray}
\tilde q_1 &=& - {{\cal L}^*(\omega)\over{\cal L}(\omega)}\tilde{p}_1
+ \frac{i \omega \sqrt{\delta \delta_{\rm e}} }{ {\cal L}(\omega)}
	(\tilde{n}_{e1}-\tilde{n}_{n1}) 
+ \frac{\Omega  \sqrt{2\delta \delta_{\rm e}}}{ {\cal L}
	(\omega)} \tilde{s}_1 \nonumber \\
&& \quad
- \frac{\sqrt{T_{\rm e}} 
	(\Omega^2 - \omega^2 + i \omega \delta_{\rm s})}{{\cal L}(\omega)}
	\widetilde{w}_1 
-\frac{i \omega \sqrt{2 \delta \delta_{\epsilon}} }{ {\cal L}(\omega)} 
	\widetilde{m}_{s1} \nonumber \\
&& \quad 
+\frac{\omega \sqrt{2 L \delta \delta_\epsilon} (\omega-i \delta_{\rm i})}
{\sqrt{c\delta_{\rm i}}{\cal L}(\omega)} \widetilde{m}_{n1}
\,,
\label{q1outlossy} \\
\tilde{q}_2 &=&
   \frac{2 i \omega \sqrt{ \omega_0 T_{\rm o} W_{\rm circ}^\ast}}
   	{L \sqrt{\hbar} {\cal L}(\omega)} \tilde{x}
-\frac{{\cal L}^\ast (\omega)}{{\cal L}(\omega)}  \tilde{p}_2
+\frac{ \Omega \sqrt{2\delta \delta_{\rm e}}}{ {\cal L}(\omega)}
	\tilde{s}_2 \nonumber \\
&& \quad 
+\frac{i \omega \sqrt{\delta \delta_{\rm e}}}{ {\cal L}(\omega)}
	(\tilde{n}_{e2}-\tilde{n}_{n2})  
- \frac{\sqrt{T_{\rm e}} (\Omega^2 - \omega^2 + i \omega \delta_{\rm s} ) }
	{ {\cal L}(\omega)}\widetilde{w}_2 \nonumber \\
&& \quad 
-\frac{i \omega \sqrt{2 \delta \delta_{\epsilon}} }{ {\cal L}(\omega)} 
	\widetilde{m}_{s2} 
+\frac{\omega \sqrt{2 L \delta \delta_\epsilon} (\omega-i \delta_{\rm i})}
{\sqrt{c\delta_{\rm i}}{\cal L}(\omega)} \widetilde{m}_{n2}
\,,
\label{q2outlossy}
\end{eqnarray}
\label{qsoutlossy}
\end{mathletters}
where 
\begin{eqnarray}
\tilde{x} &=& L \tilde{h} -
\frac{4 \sqrt{2 \hbar \omega_0 W_{\rm circ}^\ast}}{mc \omega^2 {\cal L}(\omega)}
\Biggl[ 
	\frac{i \omega \sqrt{2c \delta}}{\sqrt{L}} \tilde{p}_1
	+\frac{i \omega \sqrt{c \delta_\epsilon}}{\sqrt{L}} \widetilde{m}_{s1}
	\nonumber \\
	&& \quad \qquad \qquad
	-\frac{\sqrt{\delta_\epsilon}[\Omega^2-i\omega (\delta+\delta_{\rm i})]}
		{\sqrt{\delta_{\rm i}}} \widetilde{m}_{n1}
	-\frac{\Omega \sqrt{c \delta_{\rm e}}}{ \sqrt{L}} \tilde{s}_1
	\nonumber \\
	&& \quad \qquad \qquad
	+i \omega \sqrt{\delta \delta_{\rm e}} \widetilde{w}_1
	-\frac{i \omega \sqrt{c \delta_{\rm e}}}{\sqrt{2L}}
		(\tilde{n}_{e1}-\tilde{n}_{n1})
   	  \Biggr] 
\label{lossyxba}
\end{eqnarray}
with 
\bea
\delta_{\rm e} &= c T_{\rm e}/2L \,, \quad
\delta_{\rm s} &= c T_{\rm s}/2L \,, \nonumber \\
\delta_{\rm i} &= c T_{\rm i}/4L \,, \quad
\delta_\epsilon &= c {\cal E}/2L \,.  
\eea
Note that the 
expression for the circulating power now has the form
\begin{equation}
W_{\rm circ}^\ast = \frac{1}{2} \hbar \omega_0 B_1^2 
	= \frac{4 \hbar \omega_0 T_{\rm i} T_{\rm p} I_1^2}
			{(T_{\rm i} T_{\rm p}+4T_{\rm e})^2} 
\end{equation}
[cf. Eq.~(\ref{Wcirc})].

Equations (\ref{qsoutlossy}) are approximate expressions [accurate to
about 6\%, as were Eqs.~(\ref{qsout}); see Footnote~\ref{note:output}], where
the assumptions (\ref{newtransreq}) regarding the relative sizes of the
transmissivities were used to simplify from the exact expressions.  
Alternatively, they can be derived 
analytically by keeping the leading order of
the small quantities 
\( \omega L/c\sim \sqrt{T_{\mathrm{s}}}\sim T_{\mathrm{o}}\sim T_{\mathrm{i}} \),
plus the various loss factors; see Sec.~VI of KLMTV and 
%Sec.~\ref{sec:lossy} of this thesis
Sec.~IV of Paper~I 
for details of the derivations for other inteferometer designs.
In addition to confirming the approximate
formulas, such a derivation can also clarify the origins of various noise
terms and their connections to one another. 

\subsection{Internal and External Losses in Compact Form}

In order to simplify the above Eqs.~(\ref{qsoutlossy}) and 
(\ref{lossyxba}), we define $\kappa^*$ in identically the same way
as we defined $\kappa$ [Eq.~(\ref{kappa}) or (\ref{kappasimple})] but with
$W_{\rm circ} \rightarrow W_{\rm circ}^*$.  Let 
\( \mathcal{E}^{\mathrm{S}}_{\mathcal{N}} \)
and \( \mathcal{E}^{\mathrm{R}}_{\mathcal{N}} \) represent the shot 
and radiation-pressure noises for the various parts of the 
interferometer, specified by $\cal N$. In Table \ref{table:lossfactor},
expressions for \( \mathcal{E}^{\mathrm{S}}_{\mathcal{N}} \)
and \( \mathcal{E}^{\mathrm{R}}_{\mathcal{N}} \) are given for 
$\cal N =$ AES (arm cavities, extraction mirror, and sloshing cavity
combined), close (port-closing mirror), $\rm RSE_{in}$
(RSE cavity in the north direction, or going ``in" to the arms), 
and $\rm RSE_{out}$ (RSE cavity in the south direction, 
or going ``out" of the arms).  The various $\varepsilon_{\cal N}$
represent the characteristic (and frequency-independent) 
fractional losses for each of these terms; values are given in 
Table \ref{table:fraclosses}.
Note that, by definition, \( \mathcal{E}^{\mathrm{S}}_{\mathcal{N}} \)
are required to be real, while \( \mathcal{E}^{\mathrm{R}}_{\mathcal{N}} \)
may have imaginary parts.  For more information, including physical 
explanations of each of these terms, see Appendix \ref{app:losses}.

It is simple at this point to include the losses associated with
optical elements external to the interferometer.  These include losses
are associated with 
\begin{itemize}
\item the local oscillator used for homodyne detection,
\item the inefficiency of the photodiode, 
\item the circulator by which the squeezed vacuum is injected, and 
\item the external filter cavities used for the variational-output scheme.
\end{itemize}
These can be addressed in the same manner as the losses inside the speed meter.
We need only include two more terms in the summation,  
\( \mathcal{N}=\mathrm{OPC} \) for the local oscillator, photodiode,
and circulator and \( \mathcal{N}=\mathrm{F} \)
for the filters. 
Again, these terms are shown in Tables~\ref{table:lossfactor} and 
\ref{table:fraclosses}
and described in more
detail in Appendix \ref{app:losses}. 

{\setlength{\tabcolsep}{0.185cm}
\begin{table*}[t]
\caption[Loss factors due to shot noise and radiation pressure for each 
type of cavity in the interferometer]
{Loss factors $\cal E_N^{\rm S}$ due to shot noise and 
$\cal E_N^{\rm R}$ due to radiation pressure for each 
type of loss source in the interferometer. \label{table:lossfactor}}
\begin{tabular}{lcccccc}
%\hline
%\hline 
Source&${\cal N}$ & \hspace*{0in} &
\( \mathcal{E}^{\mathrm{S}}_{\cal N} \) (shot noise) & \hspace*{0in} &
\( \mathcal{E}^{\mathrm{R}}_{\cal N} \) (radiation pressure noise) &  \\
%\hline
\hline 
%\hspace*{-.09in}
$\begin{array}{l}
 {\rm arm\ cavities,} \\
  {\rm extract.\ mirror,} \\
{\rm sloshing\ cavity}
 \end{array}$ &\( \mathrm{AES} \)&&
\( \displaystyle \sqrt{\frac{\varepsilon _{\mathrm{AES}}}{T_{\mathrm{o}}}}
	\frac{\omega \delta }{|\mathcal{L}(\omega )|} \)&&
\( \displaystyle -\frac{e^{i\psi }}{2}
	\sqrt{\frac{\varepsilon _{\mathrm{AES}}}{{T}_{\rm o}}} \) &\\
%\hline 
&&&&& \\
%\hspace*{-.09in}
$\begin{array}{l}
{\rm \textrm{port-closing}} \\
{\rm mirror}
 \end{array}$
&\( \mathrm{close} \)&&
\( \displaystyle \sqrt{\varepsilon _{\mathrm{close}}} \,
	\frac{\Omega ^{2}-\omega ^{2}}{|\mathcal{L}(\omega )|} \)&&
\( \displaystyle -\frac{ie^{i\psi }}{2}
	\sqrt{\varepsilon _{\mathrm{close}}} \)&\\
%\hline 
&&&&& \\
%\hspace*{-.09in}
$\begin{array}{l}
{\rm RSE\ cavity}\\
{\rm ``in"\ to\ arms} 
\end{array}$&\( \mathrm{RSE}_{\mathrm{in}} \)&&
\( \displaystyle 
	\sqrt{\frac{\varepsilon _{\mathrm{RSE}}T_{\mathrm{i}}}{4T_{\mathrm{o}}}
	\left( 1+\frac{\omega ^{2}}{\delta _{\mathrm{i}}^{2}}\right) }
	\frac{\omega \delta }{|\mathcal{L}(\omega )|} \)&&
\( \displaystyle e^{i\psi -i\beta _{\mathrm{i}}}
	\sqrt{\frac{\varepsilon _{\mathrm{RSE}}T_{\mathrm{o}}}{T_{\rm i}}}
	\frac{\omega (\delta _{\mathrm{i}}+\delta )+i\Omega ^{2}}{\omega \delta } \)&\\
%\hline 
&&&&& \\
%\hspace*{-.09in}
$\begin{array}{l}{\rm RSE\ cavity} \\
	{\rm ``out"\ to\ slosh}\end{array}$&\( \mathrm{RSE}_{\mathrm{out}} \)&&
\( \displaystyle 
	\sqrt{\frac{\varepsilon _{\mathrm{RSE}}T_{\mathrm{i}}}{4T_{\mathrm{o}}}
	\left( 1+\frac{\omega ^{2}}{\delta _{\mathrm{i}}^{2}}\right) }
	\frac{\omega \delta }{|\mathcal{L}(\omega )|} \)&&
\( \displaystyle e^{i\psi +i\beta _{\mathrm{i}}}
	\sqrt{\frac{\varepsilon _{\mathrm{RSE}}T_{\mathrm{o}}}{T_{\rm i}}}
	\frac{\omega (\delta _{\mathrm{i}}-\delta )-i\Omega ^{2}}{\omega \delta } \)&\\
%\hline
&&&&& \\
%\hspace*{-.09in}
$\begin{array}{l}
{\rm local\ oscillator,} \\
{\rm photodiode,} \\
{\rm and\ circulator} \end{array}$
&OPC &&
$\displaystyle \sqrt{\varepsilon_{\rm OPC}}$ && 0 & \\
&&&&& \\
filter cavities&F &&
%$\displaystyle \frac{1}{2} 
%\sum_{\stackrel{J=\rm I,II}{s=+,-}}
%	\frac{4 T_e}{ T_{J}
%		\left[1+(s \omega/\delta_{J}-\xi_{J})^2\right]}$
$\sqrt{\varepsilon_{\rm F}}$
	  && 0 & \\
&&&&& \\
%\hline
%\hline
\end{tabular}
\end{table*}}

%{\setlength{\tabcolsep}{0.18cm}
\begin{table}
\caption[Fiducial values for the fractional losses]
{Fiducial values for the fractional losses occuring in various parts
 of the interferometer.  These losses and their values are discussed in
 more detail in Appendix \ref{app:losses}.\label{table:fraclosses}}
%\begin{minipage}[t]{6in}
\begin{tabular}{lllll}
%\hline
%\hline
Loss source & Symbol & Value \\
\hline
arm cavity & $\varepsilon_{\rm arm}$ & $2 \times 10^{-5}$ \\
sloshing cavity &$\varepsilon_{\rm slosh}$ & $2 \times 10^{-5}$ \\
extraction mirror & $\varepsilon_{\rm ext}$ & $2 \times 10^{-5}$ \\
RSE cavity & $\varepsilon_{\rm RSE}$ & $2 \times 10^{-5}$ \\
port-closing mirror & $\varepsilon_{\rm close}$ & $2 \times 10^{-5}$ \\
local oscillator & $\varepsilon_{\rm lo}$ & $0.001$ \\
photodiode & $\varepsilon_{\rm pd}$ & $0.001$ \\
circulator & $\varepsilon_{\rm circ}$ & $0.001$ \\
mode-mismatch into filters & $\varepsilon_{\rm mm}$ & $0.001$ \\
\hline
Combined loss source terms && \\
\hline
arms, extraction mirror, \& sloshing cavity\tablenote{This 
loss does have some weak frequency dependence, shown in 
Eq.~(\ref{epAES}), which will cause it to increase slightly at very low
frequencies.}
	&$\varepsilon_{\rm AES}$ & $6 \times 10^{-5}$ \\
local oscillator, photodiode, \& circulator & $\varepsilon_{\rm OPC}$ & $0.003$ \\
filter cavities (with mode mismatch) & $\varepsilon_{\rm F}$ & $0.005$ \\
\end{tabular}
%\end{minipage}
%\begin{tabular}{lllll}
%\hline
%\hline
%\hspace*{1in} & \hspace*{1in} & \hspace*{1in} & \hspace*{1in} & \hspace*{1.16in} \\
%\end{tabular}
%\vspace*{-2mm}
\end{table}
%}

Using these $\cal E_N^{\rm S}$ and $\cal E_N^{\rm R}$,
we can rewrite the input-output relation (\ref{qsoutlossy}) in 
the same form as Eq.~(\ref{qsoutKimbleform}) as follows:
\bea
\label{inoutloss}
\left( \begin{array}{c}
\tilde{q}_{1}\\
\tilde{q}_{2}
\end{array}\right) &=& e^{2i\psi }\left( \begin{array}{cc}
1 & 0\\
-\kappa ^{*} & 1
\end{array}\right) \left( \begin{array}{c}
\tilde{p}_{1}\\
\tilde{p}_{2}
\end{array}\right) 
\nonumber \\
&& \quad 
+\sum _{\mathcal{N}}e^{2i\alpha _{\mathcal{N}}}\left( \begin{array}{rr}
\mathcal{E}^{S}_{\mathcal{N}} & 0\\
-\kappa ^{*}\mathcal{E}^{\mathrm{R}}_{\mathcal{N}} & \mathcal{E}^{S}_{\mathcal{N}}
\end{array}\right) \left( \begin{array}{c}
n_{{\cal N}1}\\
n_{{\cal N}2}
\end{array}\right) 
\nonumber \\
&& \quad 
+\sqrt{2\kappa ^{*}}\frac{h}{h_{{SQL}}}e^{i\psi }\left( \begin{array}{c}
0\\
1
\end{array}\right) \, ,
\eea
where the \( \alpha _{\mathcal{N}} \) are uninteresting
phases that do not affect the noise.

\begin{figure}[t]
\epsfig{file=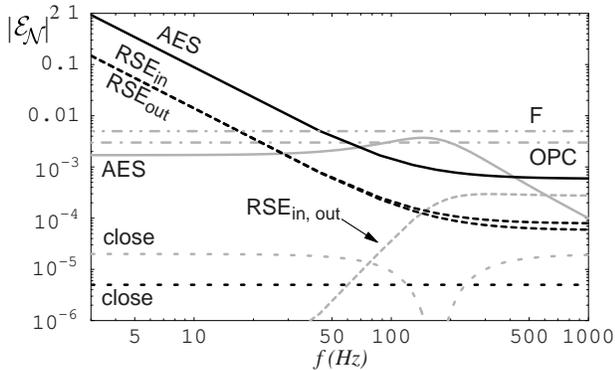,width=0.475\textwidth} 
\caption[Moduli-squared of the loss factors]{\label{lossfactorplot}
Moduli-squared of the loss factors shown in Table \ref{table:lossfactor}.
In general, the black curves are the radiation-pressure noise and 
the gray curves are the shot noise.
The parameters used for this
plot are given in Tables \ref{table:newparams} and \ref{table:fraclosses}.
}
\label{fig:lossfactors}
\end{figure}

The relative magnitudes of the loss terms are shown 
in Fig.~\ref{fig:lossfactors}. 
From the plot, we can see that there are several
loss terms---specifically, the shot noise from the AES, OPC, and
filter cavities (if any)---that are of comparable magnitude at high
frequencies and dominate there.  The AES radiation-pressure term dominates
at low frequencies, and the RSE radiation-pressure terms are also
significant.  Since the largest noise sources at low 
frequencies are radiation-pressure terms, they will be dependent
on the circulating power.  Consequently, those terms will become
smaller when the circulating power is reduced, as when squeezed vacuum is
injected into the dark port.  This will be demonstrated in 
Fig.~\ref{fig:lossynoise} below.

To compute the spectral noise density, we 
suppose the output at homodyne angle \( \Phi  \) is measured, giving
\bea
\label{totalnoiseloss}
S_{h_n} (\omega ) &=& \frac{(h_{{SQL}})^2}{2\kappa ^{*}}
	\Biggl\{ \bigg[(\cot \Phi -\kappa ^{*})^{2}+1\bigg ]
\nonumber \\
&& \quad
		+\sum _{\mathcal{N}}\bigg [|{\cal E}^{\mathrm{S}}_{\mathcal{N}}\cot \Phi 
			-{\cal E}^{\mathrm{R}}_{\mathcal{N}}\kappa^{*}|^2
			+({\cal E}^{\mathrm{S}}_{\mathcal{N}})^2 \bigg ]
	\Biggr\} \, ,
\eea
where we have assumed all of the vacuum fluctuation spectral densities 
are unity and the cross-correlations are zero; this is the same technique
that we used to derive Eqs.~(\ref{simplespec}) and (\ref{shns}) and that was used
in 
%Chapter 2
Paper~I
and KLMTV.
Given the complicated behaviors of \( \mathcal{E}^{\mathrm{S}}_{\cal N} \) 
and \( \mathcal{E}^{\mathrm{R}}_{\cal N} \),
including these loss terms in the optimization of the homodyne phase
\( \Phi (\omega)  \)
is unlikely to be helpful. Therefore, we will use 
\( \cot \Phi =\kappa _{\mathrm{max}}^{*} \), as in the lossless case.
This gives us a total noise with losses:
\bea
\label{SlossSimpleOpt}
S_{h_n} (\omega ) &=& \frac{(h_{{SQL}})^2}{2\kappa ^{*}}
	\Biggl\{ \bigg[ (\kappa _{\mathrm{max}}^{*}-\kappa ^{*})^{2}+1\bigg ]
\nonumber \\
&& \quad
		+\sum _{\mathcal{N}}\bigg[ |{\cal E}^{\mathrm{S}}_{\mathcal{N}}
			\kappa ^{*}_{\mathrm{max}}
		-{\cal E}^{\mathrm{R}}_{\mathcal{N}}\kappa^{*}|^{2}
		+({\cal E}^{\mathrm{S}}_{\mathcal{N}})^{2}\bigg ]
	\Biggr\} \,.
\eea

When we inject squeezed vacuum into the dark port, we get output operators
\bea
\label{inoutlosssqueeze}
\left( \begin{array}{c}
\tilde{q}_{1s}\\
\tilde{q}_{2s}
\end{array}\right) &=&
e^{2i\psi }\left( \begin{array}{cc}
1 & 0\\
-\kappa ^{*} & 1
\end{array}\right) \left( \begin{array}{c}
e^{R}\tilde{p}_{1}\\
e^{-R}\tilde{p}_{2}
\end{array}\right)
\nonumber \\
&& \quad 
+\sum _{\mathcal{N}}e^{2i\alpha _{\mathcal{N}}}
\left( \begin{array}{rr}
\mathcal{E}^{S}_{\mathcal{N}} & 0\\
-\kappa ^{*}\mathcal{E}^{\mathrm{R}}_{\mathcal{N}} & 
	\mathcal{E}^{S}_{\mathcal{N}}
\end{array}\right) 
\left( \begin{array}{c}
n_{{\cal N}1}\\
n_{{\cal N}2}
\end{array}\right) \nonumber \\
&& \quad
+\sqrt{2\kappa ^{*}}\frac{h}{h_{{SQL}}}e^{i\psi }
\left( \begin{array}{c}
0\\
1
\end{array}\right) 
\eea
that can be regarded as acting on the ordinary vacuum states of the input.
Once again assuming that the vacuum fluctuation spectral densities 
are unity and the cross-correlations are zero,
the squeezed-input noise spectral density with homodyne detection
at phase \( \Phi  \) is
\bea
\label{SlossSimpleOptSqueeze}
S_{h_{ns}}(\omega )
& = & \frac{(h_{{SQL}})^2}{2\kappa ^{*}}
	\Biggl\{ \bigg [(\cot \Phi -\kappa ^{*})^{2}e^{2R}+e^{-2R}\bigg ]
\nonumber \\
&& \
	+\sum _{\mathcal{N}}\bigg [|{\cal E}^{\mathrm{S}}_{\mathcal{N}}\cot \Phi 
		-{\cal E}^{\mathrm{R}}_{\mathcal{N}}\kappa ^{*}|^{2}
		+({\cal E}^{\mathrm{S}}_{\mathcal{N}})^{2}\bigg ]
	\Biggr\} \, .
\eea

\subsection{Performance of Lossy Speed Meters and Comparisons with
  Other Configurations}
\label{sec:compare}

\begin{figure}[t]
\epsfig{file=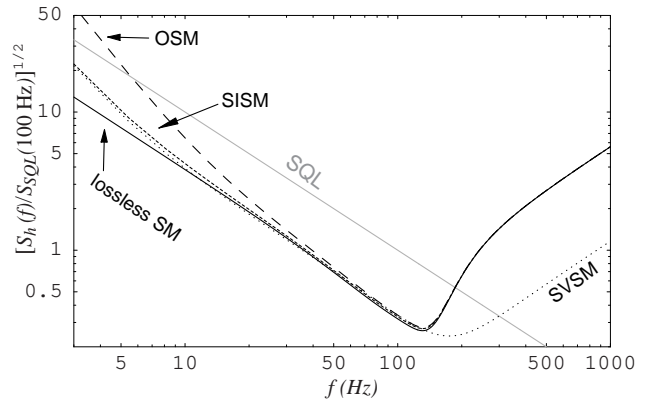,width=0.475\textwidth} 
\caption[Noise curves showing the effects of losses]
{Noise curves showing the effects of losses.  Noise curves for lossy 
versions of the
ordinary (OSM), squeezed-input (SISM), and squeezed-variational (SVSM)
speed meters are shown, along with a curve of the lossless ordinary
speed meter for comparison.   All speed 
meter curves here have the same parameters: $\delta = 2 \omega_{\rm opt}$,
$\Omega=\sqrt{3}\omega_{\rm opt}$, $\omega_{\rm opt}=2\pi \times 100~\rm Hz$,
and $T_{\rm i}=0.005$. The rest of the parameters are given in Tables
\ref{table:newparams} and \ref{table:fraclosses}. 
\label{fig:lossynoise}}
\end{figure}

Examples of lossy speed meter noise curves with and without squeezed vacuum 
[Eqs.~(\ref{SlossSimpleOpt}) and (\ref{SlossSimpleOptSqueeze})]
are shown
in Fig.~\ref{fig:lossynoise}. 
%[Parameters in 
%Tables \ref{table:newparams} and \ref{table:fraclosses} are
%used.]  
Note that, as mentioned before, the losses are less significant
when squeezed vacuum is used to reduce the circulating power, 
since the radiation-pressure noise coming from the losses is
reduced.  In the ordinary speed meter (no squeezed vacuum), the losses increase
$\sqrt{S_{h_n}}$ by $5-9\%$ in the band $50-105~\rm Hz$.  The losses
have little effect above this range, but below it, noise increases 
significantly, mostly due to the radiation-pressure noises shown in
Fig.~\ref{fig:lossfactors}.  For the squeezed-input speed meter
(power squeeze-factor $e^{-2R}=0.1$), the losses increase
$\sqrt{S_{h_n}}$ by $3-4\%$ in the band $25-150~\rm Hz$. 
Again, the losses
have little effect above this range.  At low frequencies, however,
the losses get quite large: \( 11\% \) at 10\,Hz, \( 32\% \) at 5\,Hz, 
and \( 73\% \) at 3\,Hz.  Losses in the squeezed-variational speed meter 
are
much the same as in the squeezed-input speed meter.
The slight difference at low frequencies is due to the fact that the
lossless squeezed-variational speed meter is slightly better in that regime
than the ordinary or squeezed-input speed meter.

\begin{figure}[t]
\epsfig{file=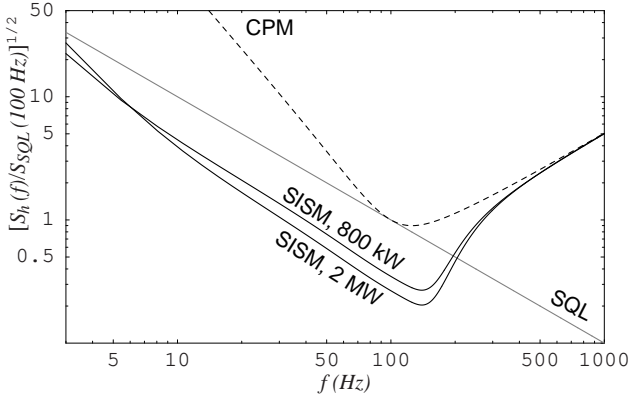,width=0.475\textwidth} 
\caption[Comparison of noise curves of a conventional interferometer
and a speed meter]{Comparison of noise curves of a conventional position meter
(CPM) and squeezed-input speed meters (SISM) with circulating 
powers $W_{\rm circ}= 820~\rm kW$ and $W_{\rm circ}= 2~\rm MW$.
The speed meters have $f_{\rm opt}=107~\rm Hz$, with $\Omega$ and $\delta$ 
determined by Eq.~(\ref{doublexi}).
Other parameters used are those in Tables~\ref{table:newparams} and 
\ref{table:fraclosses} with $T_{\rm i}=0.005$ and $e^{-2R}=0.1$.  }
\label{fig:sqordcomp}
\end{figure}

The noise
curves of squeezed-input speed meters (with ordinary homodyne detection) compared 
with the SQL are shown in Fig.~\ref{fig:sqordcomp}, along with
the noise of a conventional position meter with the same optical
power.  These speed meters beat the SQL in a broad frequency band, despite
the losses. In particular, the noise curve for the speed meter with 
$W_{\rm circ}=800\,{\rm kW}$ (and $f_{\rm opt}=107\,{\rm Hz}$) 
matches the curve of the conventional position meter
at high frequencies, while it beats the SQL by a factor of $\sim 8$ 
(in power) below
$\sim 150\,{\rm Hz}$.  
In terms of the signal-to-noise ratio for 
neutron star binaries,
for example, this
configuration improves upon the conventional design by a factor  of
$3.6$ in signal-to-noise ratio, which corresponds to a 
factor of $43$ increase in event rate. 
If it is possible to have a higher circulating power, say 
$W_{\rm circ}=2\,{\rm MW}$, the squeezed-input speed meter
would be able to beat the SQL by a factor of $\sim 14$, corresponding
to a factor of 4.6 in signal-to-noise and 97 in event rate.  
(Such a noise curve is shown
in Fig.~\ref{fig:sqordcomp}).

The broadband behaviors of the speed meters with losses are
particularly interesting. We start by looking at the
expression for the noise spectral density, Eq.~(\ref{SlossSimpleOptSqueeze}).
An ideal (lossless) speed meter in the broadband configuration beats the SQL from
0~Hz up to $\omega \sim \omega_{\rm opt}$, by roughly a constant
factor, because $\kappa$ is roughly constant in this band. {\it This
is the essential feature of the speed meter}; see Sec.~\ref{sec:lossless}.
Focusing
on this region, we have, approximately (for squeezed-input
speed meters that are lossy):
\bea
S_{h_{ns}}(\omega) &\approx & 
	\frac{h_{SQL}^2}{2\kappa^*_{\rm max}}
		\biggl[e^{-2R}
			+\sum_{\cal N}|{\cal E}_{\cal N}^{S}|^2
\nonumber \\
&& \qquad \qquad
			+{\kappa^*}_{\rm max}^2\sum_{\cal N}
				|{\cal E}_{\cal N}^{\rm S}-{\cal E}_{\cal N}^{\rm R}|^2 
		\biggr]\,.
\eea
Qualitatively, we can see that if 
the losses are not severe or if $\kappa^*_{\rm max}$ is
relatively small (such that the later two terms in the above equation
are small compared to the power squeeze factor $e^{-2R}$), 
the losses do not contribute significantly 
to the total noise.  If, in addition, the dominant loss factors are
(almost) frequency independent, then the noise due to losses gives a rather
constant contribution, as shown by curves in
Fig.~\ref{fig:lossynoise}.  In particular, the large bandwidth is preserved.
(There is a slight exception to this statement in the absence of 
squeezed input.  Without squeezed input, the circulating power 
is higher, causing $\kappa^*_{\rm max}$ to be $10$
times larger than the other cases.  Consequently, the frequency dependence
of ${\cal E}_{\rm AES}^{\rm R}$ to appear in the output.)

\begin{figure}[t]
\epsfig{file=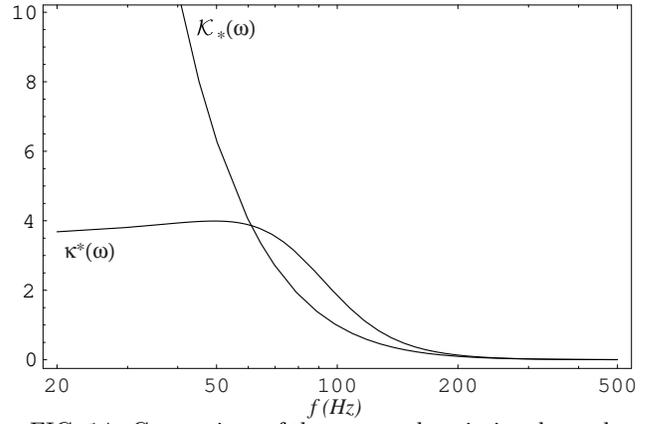,width=0.475\textwidth} 
\caption[Comparison of the speed meter's $\kappa^*$ with the position
meter's $\cal K_*$]
{Comparison of the squeezed-variational speed meter's $\kappa^*$ with the equivalent
coupling constant $\cal K_*$ (as defined by KLMTV) 
for the squeezed-variational position meter. Parameters are 
$W_{\rm circ}=820\,{\rm kW}$, 
$\gamma=\delta=2\pi\times 100\,{\rm Hz}$, $\Omega=2\pi\times 173\,{\rm Hz}$.}
\label{fig:kappacomp}
\end{figure}

\begin{figure}[t]
\epsfig{file=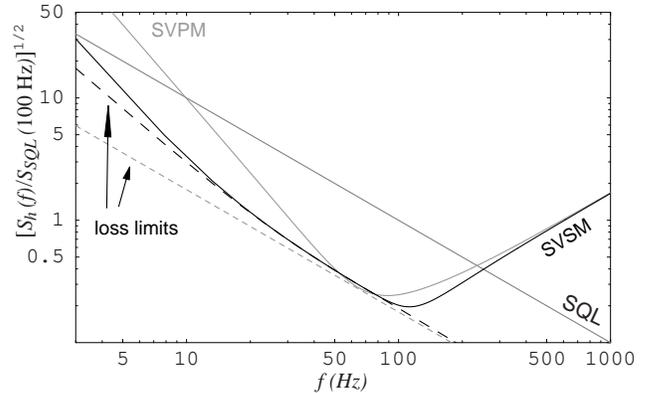,width=0.475\textwidth}  
\caption[Comparison of noise curves for a squeezed-variational
position meter and speed meter]{\label{fig:lastcompare}
Comparison of noise curves for a squeezed-variational position meter
(SVPM; analyzed in KLMTV) 
and for a squeezed-variational speed meter (SVSM; analyzed in this
paper).  Parameters used are those in Tables~\ref{table:newparams} and 
\ref{table:fraclosses} with $T_{\rm i}=0.005$ and $e^{-2R}=0.1$. 
Also shown are
the loss limits described in Sec.~\ref{sec:compare}.}
\end{figure}

As $\kappa_{\rm max}$ increases, the noise from the losses
may become dominant.  In fact, when one minimizes the noise spectral 
density with respect to $\kappa^*_{\rm max}$, one obtains the following
loss-dominated result:
\beq
S^{\rm L}_{h}(\omega) \approx  h_{SQL}^2
%\nonumber \\
%&& \; \times
\sqrt{
\left(\sum_{\cal N}|{\cal E}_{\cal N}^{\rm
    S}-{\cal E}_{\cal N}^{\rm R}|^2\right)
\left(e^{-2R}+\sum_{\cal N}|{\cal E}_{\cal
  N}^{\rm S}|^2\right)
}\,,
\eeq
which is achieved if and only if 
\beq
\kappa^*_{\rm max}=\kappa^{\rm L}
\equiv
\sqrt{
\frac
{e^{-2R}+\sum_{\cal N}|{\cal E}_{\cal
  N}^{\rm S}|^2}
{\sum_{\cal N}|{\cal E}_{\cal N}^{\rm
    S}-{\cal E}_{\cal N}^{\rm R}|^2}}\,.
\eeq
This $\kappa^{\rm L}$ is rather constant and is comparable in
magnitude to the values of $\kappa^*(\omega)$ of our
speed meters, suggesting that the speed meters can become loss-limited over
a broad band of frequencies.  Contrast this with the KLMTV position meters,
where ${\cal K}_*(\omega)$ grows as $\omega^{-2}$ at low frequencies;
see Fig.~\ref{fig:kappacomp}.  {\it This is a fundmental
property of displacement meters.}  As a result, a position meter
optimized at some frequency $f_{\rm opt}$ may be able to reach its
``loss limit" (the equivalent of $S_h^{\rm L}$) at that frequency
$f_{\rm opt}$, but doing so will result in a sharp growth of noise at
freqencies below $f_{\rm opt}$.  In contrast, a speed meter similarly
optimized is able to stay at the noise level of its loss limit $S_h^{\rm L}$
over a wide band of frequencies below $f_{\rm opt}$;
see Fig.~\ref{fig:lastcompare}.  While it is unfortunate that losses limit
the performance of interferometers, the speed meter is at least able to 
retain a wide-band sensitivity even in the presence of a loss-limit.

To give a 
specific example of this loss-limit phenomenon, we first notice that, 
with the same circulating power,
the conventional position-meter ${\cal K}_*$ and 
our (squeezed-variational) speed-meter 
$\kappa$ agree\footnote{In fact, ${\cal K}_*$ can be
obtained from the speed meter $\kappa^*$ by putting
$\Omega\rightarrow0$ and $\delta\rightarrow\gamma$.}
if $\delta=\gamma$ (where $\gamma$ is
the bandwidth of the arm cavities, as defined in KLMTV) 
and if we consider high 
frequencies ($\omega \gtrsim \{\gamma,\,\Omega\}$).  Figure~\ref{fig:kappacomp}
shows an example of this [with $W_{\rm circ}=820\,{\rm kW}$, 
$\gamma=\delta=2\pi\times 100\,{\rm Hz}$, $\Omega=2\pi\times
173\,{\rm Hz}$].  
The noise curves
of the two interferometers are shown in
Fig.~\ref{fig:lastcompare}. 
%(They also happen to have roughly the same
%loss limit, as also shown in the plot.)

As expected, the two noise curves in Fig.~\ref{fig:lastcompare} agree at very high
frequencies. At intermediate frequencies, the speed meter's 
$\kappa^*$ is larger than the position meter's ${\cal K}_*$, and 
thus the speed meter has better sensitivity than the position 
meter. As the frequency decreases, the speed meter reaches its loss limit
first and stays at that limit for a wide range of frequencies.
The position meter, however, only touches its loss limit and 
then increases rapidly.

\section{Conclusions}
\label{sec:conclusions}

We have described and analyzed a speed-meter interferometer that has
the same performance as the two-cavity design analyzed in 
%Chapter 2,
Paper~I, 
but it does so
without the substantial amount of power flowing through the system or
the exorbitantly high input laser power required by the two-cavity speed 
meter.  It was also shown that
the injection of squeezed vacuum with $e^{-2R}=0.1$ 
into the dark port of the interferometer
will reduce the needed circulating power by an order of magnitude,
bringing it into a range that is comparable to the expected circulating
power of LIGO-II, if one wishes to beat the SQL by a factor of $\sqrt{10}$
in amplitude.  Additional improvements to the sensitivity, 
particularly at high frequencies, can be achieved through the
use of frequency-dependent homodyne detection.  

In addition, it was shown that this type of speed-meter interferometer is not
nearly as susceptible to losses as those presented in KLMTV.  Its
robust performance is due, in part, to the 
functional form the coupling factor $\kappa$, which is roughly
constant at low frequencies.  This helps to maintain the speed meters'
wideband performance, even in the presence of losses.
Losses
for the various speed meters we discuss here are generally quite low.  
The dominant sources of loss-induced 
noise at low frequencies ($f \lesssim f_{\rm opt}$) 
are the radiation-pressure noise from losses in the arm, extraction,
and sloshing cavities.  Because this type of noise is dependent on
the circulating power, it can be reduced by reducing the power by means
of squeezed input.

\section*{Acknowledgments}

We thank Kip Thorne for helpful
advice about its solution and about the prose of this paper.  We also
thank Farid Khalili, Stan Whitcomb, Ken Strain, and Phil Willems 
for useful discussions.
This research was supported in part by NSF grant PHY-0099568 and the David 
and Barbara Groce Fund at the San Diego Foundation.

\appendix
\section{FP Cavities as Optical Filters}
\label{app1}

As proposed by KLMTV [Sec.~V B and Appendix~C], Fabry-Perot cavities can be
used as optical filters to achieve frequency-dependent homodyne
detection. Here we shall briefly summarize and generalize their results. 

Suppose we have
one FP cavity of length 
$L_{\rm FP}$ and resonant frequency  
$\omega_0-\xi_{\rm FP}\delta_{\rm FP}$.
Also suppose this cavity has an input mirror with
finite transmissivity $T_{\rm FP}$ and a perfect 
end mirror.  When sideband fields at frequency
$\omega_0\pm\omega$ emerge from the cavity, they have a phase shift
\beq
\label{alphapm}
\alpha_{\pm}\equiv 2 \arctan(\xi_{\rm FP}\pm\omega/\delta_{\rm FP}) \;,
\eeq
where 
\beq
\delta_{\rm FP}=\frac{c T_{\rm FP}}{4L_{\rm FP}}
\eeq
is the half bandwidth of the cavity. 
[Note that Eq.~(\ref{alphapm}) is 
KLMTV Eqs.~(88) and (C2), but a factor of 2 was missing from 
their equations.  Fortunately, this appears to be a typographical
error only in that particular equation; the factor of 2 is included in their
subsequent calculations.]
As a result of this phase shift, 
the input ($\tilde{b}_{1,2}$)--output (${b}_{1,2}$)
relation for sideband quadratures at 
frequency $\omega$ will be [KLMTV Eqs.~(78)]
\beq
\left(\ba{c} \tilde{b}_1 \\ \tilde{b}_2 \ea \right)
=
e^{i\,\alpha_{\rm m}}\,
{\bf R}_{\alpha_{\rm p}}
\left(\ba{c} b_1 \\ b_2 \ea\right)\,,
\eeq
where 
\beq
\alpha_{\rm m}\equiv \frac12 (\alpha_+-\alpha_-)\,,\quad
\alpha_{\rm p}\equiv \frac12 (\alpha_++\alpha_-)\,,
\eeq
and 
\beq
{\bf R}_{\phi}\equiv
\left(
\ba{rr} \cos \phi & -\sin\phi \\  
\sin \phi &  \cos \phi \ea 
\right)\,.
\eeq

If a frequency-independent homodyne detection at phase shift $\theta$
follows the optical filter, the measured quantity will be [KLMTV Eqs.~(81)
and (82)]
\beq
\tilde{b}_{\theta}= e^{i\alpha_{\rm m}}b_{\zeta}\,,
\eeq
where
\beq
\zeta(\omega)=\theta-\alpha_{\rm p}\equiv\theta-\frac12 (\alpha_++\alpha_-)\,.
\eeq
If more than one filter is applied in sequence (I, II, \ldots,) and 
followed 
by homodyne detection at angle $\theta$, the measured quadrature will be
[Eq.~(83)]
\beq
\label{FDhomodyne}
\zeta(\omega)=\theta-\frac 12 (\alpha_{\rm I +} + \alpha_{\rm I-}+\alpha_{\rm
II +} + \alpha_{\rm II -} +\ldots)\,.
\eeq
[Note that this $\zeta (\omega)$ (KLMTV's notation) is the same homodyne angle 
$\Phi (\omega)$ that we want to produce.]
By adjusting the parameters $\xi_J$ and $\delta_J$, one might be able
to achieve the FD homodyne phases needed. KLMTV worked out a
particular case for their design [their Sec.~V B, V C, and Appendix~C].

Here we shall seek a more complete solution that works in a large
class of situations.  With the help of Eq.~(\ref{alphapm}),
Eq.~(\ref{FDhomodyne}) can be written in an equivalent form
\bea
\label{Phipoly}
\frac{1+i\tan\zeta}{1-i\tan\zeta} 
&=&
e^{2 i\theta}
\prod_{J={\rm I},{\rm II},\ldots\,,s=\pm}
\frac
{1-i \tan \left(\alpha_{Js}/2\right)}
{1+i \tan \left(\alpha_{Js}/2\right)}\,, \nonumber \\
&=&
e^{2 i\theta}
\prod_{J={\rm I},{\rm II},\ldots\,,s=\pm}
\frac
{\omega-s(-\xi_J\delta_J-i\delta_J)}
{\omega-s(-\xi_J\delta_J+i\delta_J)}\,.
\eea 
Suppose the required $\tan\zeta(\omega)$ is a rational function in
$\omega^2$,    
\beq
\tan\zeta(\omega)=
\frac
{\sum_{k=0}^n B_k \omega^{2k}}
{\sum_{k=0}^n A_k \omega^{2k}}\,,
\eeq
where $A_{k}$ and $B_{k}$ are real constants with  $A_{n}^2+B_{n}^2>0$. 
Then Eq.~(\ref{Phipoly}) requires that, for all $\omega$, 
\bea
\label{solveparam}
&& \sum_{k=0}^n (A_k+i B_k) \omega^{2k} 
\nonumber \\
&& \quad =
D\, e^{i\theta}
\prod_{J={\rm I},{\rm II},\ldots\,,s=\pm}
\bigg[{\omega-s(-\xi_J\delta_J-i\delta_J)}\bigg]\,,
\eea
where $D$ can be any real constant. Equation (\ref{solveparam}) can be
solved as follows.  First, match the roots of the polynomials
of $\omega$ on the two sides of the equation; denote these roots
by $\pm \omega_{\rm J}$ with $J=1,2,\ldots,n$.  Then we can deduce 
that $n$ filters are
needed, and their complex resonant frequencies must be offset from 
$\omega_0$ by
\beq
\omega_J= -\delta_J \xi_J - i\delta_J \,,\quad J={\rm I},{\rm II},\ldots\,,
\eeq
where $\pm \omega_{\rm I, II,\ldots}$ [with $\Im (\omega_J) > 0$] are the $2n$ roots of 
\beq
\sum_{k=0}^n (A_k+i B_k) \,\omega^{2k}\,.
\eeq
After this, the polynomials on
the two sides of Eq.~(\ref{solveparam}) can only differ by a 
complex coefficient whose argument
determines $\theta$.  In fact, by comparing the coefficients of
$\omega^{2n}$ on both sides, we have 
\beq
\theta=\arg(A_{2n}+i B_{2n})\,.
\eeq

\section{Semi-Analytical Treatment of the Loss Terms}
\label{app:losses}

In this appendix, we present a semi-analytic treatment of each 
source of noise included in Sec.~\ref{sec:internalloss}.  We will
use a notation similar to Eq.~(\ref{qsoutKimbleform}), but in
matrix form:
\beq
\left( \begin{array}{c} 
\tilde{q}_1 \\ \tilde{q}_2 
\end{array} \right)
=\left( \begin{array}{c} 
\tilde{q}_1 \\ \tilde{q}_2 
\end{array} \right)_{\rm lossless} 
+
{\mathbf{N}_{\rm loss\ source}} \;,
\label{lossform}
\eeq
where ${\mathbf{N}_{\rm loss\ source}}$ is a vectorial representation 
of whichever source
of loss we are considering at the moment.  Each of these terms
is associated with a vacuum field of the form 
$\sqrt{\cal E (\omega)} \tilde{n}(\omega)$ 
[cf.~Eq.~(\ref{includeloss})],
which enters the interferometer and increases the level of noise present.
For generality, we let $\cal E (\omega)$ be frequency dependent.  The (constant)
characteristic fractional losses for each type of loss will be denoted
by $\varepsilon$ with an appropriate subscript.
Each loss term appearing in
Table \ref{table:lossfactor} is presented in a subsection below.

\subsection{Arms, Extraction Mirror, and Sloshing Cavity (AES)}
\label{app:aes}

The losses in the arms allow an unsqueezed vacuum field 
\( \sqrt{\varepsilon_{\mathrm{arm}}}\tilde{n}_{\mathrm{arm}} \)
to enter the optical train.  By idealizing this field as arising
entirely at the arm's end mirror, propagating the field through the
interferometer to the output port, we obtain the following contribution
to the output [cf. Eq.~(\ref{includeloss})]. The associated noise
can be put into the following form
\bea
\label{Narmloss}
{\mathbf N}_{\mathrm{arm}} &=&
	-\sqrt{\frac{\varepsilon _{\mathrm{arm}}}{T_{\mathrm{o}}}}\, 
	\Biggl[ e^{i\psi }\frac{\omega \delta }{|\mathcal{L}(\omega )|}
		\left( \begin{array}{cc}
1 & 0\\
0 & 1
\end{array}\right) 
\nonumber \\
&& \qquad
	+e^{2i\psi }\left( \begin{array}{cc}
0 & 0\\
\kappa ^{*}/2 & 0
\end{array}\right) \Biggr] \left( \begin{array}{c}
\tilde{n}_{\mathrm{arm}1}\\
\tilde{n}_{\mathrm{arm}2}
\end{array}\right) \,,
\eea
where the vacuum operators from the two arms are combined as
\beq
\tilde{n}_{\rm arm}j=\frac{\tilde{n}_{ej} - \tilde{n}_{nj}}{\sqrt{2}}\;.
\eeq
The first term (independent of $\kappa^*$) is the shot-noise contribution, 
while the second
term (proportional to $\kappa^*$) is the
radiation-pressure noise.  It turns out that several of
the other loss sources $\cal N$ have a similar mathematical form.  

We consider, specifically, the loss from the extraction mirror,
which effectively allows
\( \sqrt{\varepsilon _{\mathrm{ext}}}\,\tilde{n}_{\mathrm{ext}} \)
into the optical train.   By propagating this field through the 
interferometer to the output port, we obtain the following
contribution to the noise:
\bea
\label{Nextloss}
{\mathbf N}_{\mathrm{ext}}
&=&\sqrt{\frac{\varepsilon _{\mathrm{ext}}}{T_{\mathrm{o}}}}\, 
	\Biggl[ e^{i\psi }\frac{\omega \delta }{|\mathcal{L}(\omega )|}
		\left( \begin{array}{cc}
1 & 0\\
0 & 1
\end{array}\right) 
\nonumber \\
&& \qquad \qquad
+e^{2i\psi }\left( \begin{array}{cc}
0 & 0\\
\kappa ^{*}/2 & 0
\end{array}\right) \Biggr] \left( \begin{array}{c}
\tilde{n}_{\mathrm{ext}1}\\
\tilde{n}_{\mathrm{ext}2}
\end{array}\right) \, .
\eea

The loss from the sloshing cavity is a bit different: the imperfect
end mirror of the sloshing cavity produces a vacuum noise field
$\sqrt{\varepsilon _{\mathrm{slosh}}}\,\tilde{n}_{\mathrm{slosh}}$
which exits
the cavity with the form
\begin{equation}
\label{coeffslosh}
\sqrt{\frac{4\varepsilon _{\mathrm{slosh}}/T_{\mathrm{s}}}
	{1+\omega ^{2}/(\delta _{\mathrm{s}}/2)^{2}}}
	e^{i\beta_{\mathrm{s}}}\tilde{n}_{\mathrm{slosh}\, 1,2}
\approx \sqrt{\varepsilon _{\mathrm{slosh}}}\frac{i\Omega }{\omega }
\tilde{n}_{\mathrm{slosh}\, 1,2}\, ,
\end{equation}
where  
\( \beta_{\mathrm{s}}\equiv \arctan (2\omega /\delta _{s})\approx \pi /2 \)
for most of the frequency band of interest. The associated
noise is
\bea
\label{Nsloshloss}
{\mathbf N}_{\mathrm{slosh}}
&=&-\sqrt{\frac{\varepsilon _{\mathrm{slosh}}}{T_{\mathrm{o}}}}
	\frac{i\Omega }{\omega }\, 
	\Biggl[ e^{i\psi }\frac{\omega \delta }{|\mathcal{L}(\omega )|}
	\left( \begin{array}{cc}
1 & 0\\
0 & 1
\end{array}\right) 
\nonumber \\
&& \qquad \qquad
+e^{2i\psi }\left( \begin{array}{cc}
0 & 0\\
\kappa ^{*}/2 & 0
\end{array}\right) \Biggr] \left( \begin{array}{c}
\tilde{n}_{\mathrm{slosh}1}\\
\tilde{n}_{\mathrm{slosh}2}
\end{array}\right) \, .
\eea
 
Since the vacuum fields \( \tilde{n}_{\mathrm{arm}} \), 
\( \tilde{n}_{\mathrm{ext}} \),
and \( \tilde{n}_{\mathrm{slosh}} \) are independent and uncorrelated, 
we can effectively
combine these four noises into a single expression
\bea
\label{NAESloss}
{\mathbf N}_{\mathrm{AES}} &=&
\sqrt{\frac{\varepsilon _{\mathrm{AES}}}{T_{\mathrm{o}}}}\, 
	\Biggl[ e^{i\psi }\frac{\omega \delta }{|\mathcal{L}(\omega )|}
	\left( \begin{array}{cc}
1 & 0\\
0 & 1
\end{array}\right) 
\nonumber \\
&& \qquad \qquad
+e^{2i\psi }\left( \begin{array}{cc}
0 & 0\\
\kappa ^{*}/2 & 0
\end{array}\right) \Biggr] \left( \begin{array}{c}
\tilde{n}_{\mathrm{AES}1}\\
\tilde{n}_{\mathrm{AES}2}
\end{array}\right) \, ,
\eea
with
\beq
\varepsilon _{\mathrm{AES}} \sim {\cal E}_{\rm AES}(\omega) 
\equiv \varepsilon _{\mathrm{arm}}
	+\varepsilon _{\mathrm{ext}}
	+\varepsilon _{\mathrm{slosh}}\Omega ^{2}/\omega ^{2}\;.
\label{epAES}
\eeq
We expect that $\varepsilon_{\rm arm} \sim \varepsilon_{\rm slosh} \sim
\varepsilon_{\rm ext} \sim 2 \times 10^{-5}$, as discussed in the 
paragraph following
Eq.~(\ref{includeloss}) and as shown in Table~\ref{table:fraclosses}.

\subsection{Port-Closing Mirror}

The imperfection of the closing mirror has two effects: (i) it introduces
directly a fluctuation 
\( -\sqrt{\varepsilon _{\mathrm{close}}R_{\mathrm{o}}}\,\tilde{n}_{\mathrm{close}} \)
into the output, giving a shot noise 
\begin{equation}
\label{closedirect}
{\mathbf N}^{\mathrm{shot}\, \mathrm{direct}}_{\mathrm{close}}
	=-\sqrt{\varepsilon _{\mathrm{close}}R_{\mathrm{o}}}
	\left( \begin{array}{c}
\tilde{n}_{\mathrm{close}1}\\
\tilde{n}_{\mathrm{close}2}
\end{array}\right) \, ;
\end{equation}
and (ii) it introduces a fluctuation 
\( \sqrt{\varepsilon _{\mathrm{close}}T_{\mathrm{o}}}\,\tilde{n}_{\mathrm{close}} \)
into the light that passes from the arms into the sloshing cavity,
giving (after propagation through the sloshing cavity and interferometer
and into the output):
\bea
\label{closeOT}
{\mathbf N}^{\mathrm{indirect}}_{\mathrm{close}}
	&=&-\sqrt{\varepsilon _{\mathrm{close}}}\, 
	\Biggl[ e^{i\psi }\frac{\omega \delta }{|\mathcal{L}(\omega )|}
	\left( \begin{array}{cc}
1 & 0\\
0 & 1
\end{array}\right) 
\nonumber \\
&& \qquad \quad
+e^{2i\psi }\left( \begin{array}{cc}
0 & 0\\
\kappa ^{*}/2 & 0
\end{array}\right) \Biggr] \left( \begin{array}{c}
\tilde{n}_{\mathrm{close}1}\\
\tilde{n}_{\mathrm{close}2}
\end{array}\right) \, .
\eea
Combining these two expressions gives, to leading order (in the various
transmissivities and the small parameters $\omega L/c$ and 
$\varepsilon_{\rm close}$),
\bea
\label{closelossfull}
{\mathbf N}_{\mathrm{close}}
	&=&\sqrt{\varepsilon _{\mathrm{close}}}\, 
	\Biggl[ ie^{i\psi }\frac{\Omega ^{2}-\omega ^{2}}{|\mathcal{L}(\omega )|}
	\left( \begin{array}{cc}
1 & 0\\
0 & 1
\end{array}\right) 
\nonumber \\
&& \qquad \quad
-e^{2i\psi }\left( \begin{array}{cc}
0 & 0\\
\kappa ^{*}/2 & 0
\end{array}\right) \Biggr] \left( \begin{array}{c}
\tilde{n}_{\mathrm{close}1}\\
\tilde{n}_{\mathrm{close}2}
\end{array}\right) \, .
\eea
Since $\varepsilon_{\rm close}$ is simply the loss from the port-closing
mirror itself, we can assume that 
$\varepsilon_{\rm close} \lesssim 2 \times 10^{-5}$.
Then, this and the above expression (\ref{closelossfull})
show that the output noise from the closing mirror 
is \( T_{\mathrm{o}} \) times smaller than the AES loss
[Eq.~(\ref{epAES})].

\subsection{The RSE Cavity}
\label{app:RSEcavity}

The losses in the region between the internal mirrors and the RSE
mirror, i.e., the RSE cavity, are more complicated than the previous
cases. 
As before, we suppose that, during each propagation from one end to the
other of the RSE cavity, a fraction \( \varepsilon _{\mathrm{RSE}} \) of the
light power is dissipated and replaced by a corresponding vacuum
field, \( \sqrt{\varepsilon_{\rm RSE}}\, \tilde{n}_{\mathrm{in}} \) 
or \( \sqrt{\varepsilon_{\rm RSE}}\, \tilde{n}_{\mathrm{out}} \) (depending
whether the light is propagating in towards the arms or out towards
the extraction mirror and sloshing cavity).  These
two fields \( \tilde{n}_{\mathrm{in}} \)
and \( \tilde{n}_{\mathrm{out}} \) are independent vacuum fields.  At the
leading order in \( \varepsilon _{\mathrm{RSE}} \), we have a modified version
of the ``input--output" relation for the RSE cavity:
\bea
\label{lossRSE}
\left( \begin{array}{c}
B\\
D
\end{array}\right) &=&
	\left( \begin{array}{cc}
1-\frac{1+R_{\mathrm{i}}}{2T_{\mathrm{i}}}\varepsilon _{\mathrm{RSE}} & \frac{\sqrt{R_{\mathrm{i}}}}{\mathrm{T}_{\mathrm{i}}}\varepsilon _{\mathrm{RSE}}\\
\frac{\sqrt{R_{\mathrm{i}}}}{\mathrm{T}_{\mathrm{i}}}\varepsilon _{\mathrm{RSE}} & 1-\frac{1+R_{\mathrm{i}}}{2T_{\mathrm{i}}}\varepsilon _{\mathrm{RSE}}
\end{array}\right) \left( \begin{array}{c}
A\\
C
\end{array}\right) 
\nonumber \\
&& \,
+\sqrt{\frac{\varepsilon _{\mathrm{RSE}}}{T_{\mathrm{i}}}}
\left( \begin{array}{cc}
1 & -\sqrt{R_{\mathrm{i}}}\\
-\sqrt{R_{\mathrm{i}}} & 1
\end{array}\right) \left( \begin{array}{c}
\tilde{n}_{\mathrm{in}}\\
\tilde{n}_{\mathrm{out}}
\end{array}\right) \,,
\eea
where $A,B,C,D$ are 
%the contributions of $\tilde{n}_{\rm in}$ and $\tilde{n}_{\rm out}$ to 
the field amplitudes shown in Fig.~\ref{fig:oneport3IM}.
Note that, for simplicity, we are looking at only one arm; we could equally
well use the other (substituting $B\rightarrow F$ and $C\rightarrow G$)
or the proper combination of both.  Also, notice that if 
$\varepsilon _{\mathrm{RSE} }=0$, then we find $B=A$ and $D=C$, which
illustrates the fact that the internal and RSE mirrors have no effect on the
sidebands (described in Sec.~\ref{sec:Introduction} where we introduced
the RSE mirror).

From Eq. (\ref{lossRSE}), we find that the loss inside the RSE
cavity has two effects.
%\begin{itemize}
%\item 
First, it makes the cancellation of the effect of the internal 
and the RSE mirrors imperfect. (Recall that an RSE mirror with the
same transmissivity as the internal mirrors effectively cancels the
effect of the internal mirrors on the sidebands; this was discussed in
Sec.~\ref{sec:Introduction}.)
This imperfect cancellation will not be important 
in our situation.  Indeed, there is no corresponding 
term appearing in the input--output
relation given in Eq.~(\ref{qsoutlossy}). 
%\item 

\begin{figure}
\epsfig{file=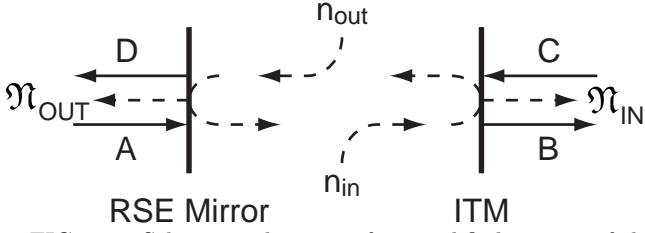,width=0.475\textwidth} 
\caption[Schematic diagram for RSE loss]{\label{fig:RSEloss}
Schematic diagram of a simplified version of the RSE cavity. 
The quantities $\tilde{n}_{\rm in}$ and $\tilde{n}_{\rm out}$ enter
inside the RSE cavity, whereas $\mathfrak{N}_{\rm in}$ and $\mathfrak{N}_{\rm out}$
are external to the cavity and exist in different locations.}
\end{figure}

Secondly, the loss inside the RSE cavity adds two vacuum 
fields to light that travels through the RSE
cavity in opposite directions [i.e., from A to B (IN) and from C to D (OUT)].
We denote them by 
\begin{mathletters}
\begin{eqnarray}
\tilde{\mathfrak N}_{\mathrm{IN}} & \equiv  & 
	\sqrt{\frac{\varepsilon_{\mathrm{RSE}}}{T_{\mathrm{i}}}}
	(\tilde{n}_{\mathrm{in}}-\sqrt{R_{\mathrm{i}}}\tilde{n}_{\mathrm{out}})\, ,\\
\tilde{\mathfrak N}_{\mathrm{OUT}} & \equiv  & 
	\sqrt{\frac{\varepsilon_{\mathrm{RSE}}}{T_{\mathrm{i}}}}
	(-\sqrt{R_{\mathrm{i}}}\tilde{n}_{\mathrm{in}}+\tilde{n}_{\mathrm{out}})\, .
\end{eqnarray}
\label{ninnout}
\end{mathletters}
Note that $\tilde{n}_{\rm in}$ and $\tilde{n}_{\rm out}$ arise 
{\it inside} the RSE cavity as a result of the loss that occurred there
and that $\tilde{\mathfrak N}_{\rm IN}$ and $\tilde{\mathfrak N}_{\rm OUT}$ 
are the vacuum fluctuations {\it emerging} from
the RSE cavity.  As a result, $\tilde{\mathfrak N}_{\rm IN}$ and 
$\tilde{\mathfrak N}_{\rm OUT}$ {\it exist in different locations:}
$\tilde{\mathfrak N}_{\rm IN}$ denotes the vacuum field {inside
the arm cavity} with $B$, and $\tilde{\mathfrak N}_{\rm OUT}$ denotes the vacuum 
field {at the RSE mirror, heading towards the extraction mirror and
sloshing cavity} with $D$.  This is depicted in Fig.~\ref{fig:RSEloss}.

%\end{itemize}

The fields $\tilde{\mathfrak N}_{\rm IN}$ and $\tilde{\mathfrak N}_{\rm OUT}$ both have a
power spectral density a factor
\( \sim 1/T_{\mathrm{i}} \) larger than the one-time loss coefficient.
This can be explained by the fact that the 
sideband light bounces back and forth inside the RSE cavity roughly
\( \sim 1/T_{\mathrm{i}} \) times before exiting.  As a result, the (power)
loss coefficient is amplified by the same factor.  However, since
these fields are quite correlated
(both contain similar amounts of \( \tilde{n}_{\mathrm{in}} \)
and \( \tilde{n}_{\mathrm{out}} \)), we need to analyze them carefully.

For the shot noise, we need to find the amplitude of the vacuum
fluctuations that the loss introduces into the output. 
To understand the effect of this type of loss, we ask how
much vacuum fluctuation is added to the field $D$ by 
\( \tilde{\mathfrak N}_{\mathrm{IN}} \) and \( \tilde{\mathfrak N}_{\mathrm{OUT}} \).
The answer is obtained by propagating \( \tilde{\mathfrak N}_{\mathrm{IN}} \) one
round trip inside the interferometer's arm(s) and then combining it with 
\( \tilde{\mathfrak N}_{\mathrm{OUT}} \).
This gives
\begin{eqnarray}
D & \rightarrow &
	 D+\left[ \tilde{\mathfrak N}_{\mathrm{OUT}}
	 	+e^{2i\omega L/c}\tilde{\mathfrak N}_{\mathrm{IN}}\right] \nonumber \\
 & \approx  & 
 	D+\sqrt{\frac{\varepsilon _{\mathrm{RSE}}T_{\mathrm{i}}}{4}
 	\left( 1+\frac{\omega ^{2}}{\delta _{\mathrm{i}}^{2}}\right) }
\nonumber \\
&& \qquad \times
 	\left( e^{i\beta _{\mathrm{i}}}\tilde{n}_{\mathrm{in}}
 		+e^{-i\beta _{\mathrm{i}}}\tilde{n}_{\mathrm{out}}\right) \, ,
\label{outflucRSE} 
\end{eqnarray}
where \( \delta _{\mathrm{i}}\equiv T_{\mathrm{i}}c/4L \) and 
\( \beta _{\mathrm{i}}\equiv \arctan (\omega /\delta _{\mathrm{i}}) \).
Propagating this to the output, we get the shot noise contribution to be
\bea
\label{NshotRSE}
{\mathbf N}^{\mathrm{shot}}_{\mathrm{RSE}}
	&=& \sqrt{\frac{\varepsilon _{\mathrm{RSE}}T_{\mathrm{i}}}{4T_{\mathrm{o}}}
	\left( 1+\frac{\omega ^{2}}{\delta _{\mathrm{i}}^{2}}\right) }
	e^{i\psi }\frac{\omega \delta }{|\mathcal{L}(\omega )|}
	\Biggl[ e^{+i\beta _{\mathrm{i}}}\left( \begin{array}{c}
\tilde{n}_{\mathrm{in}1}\\
\tilde{n}_{\mathrm{in}2}
\end{array}\right) 
\nonumber \\
&& \qquad \qquad
+e^{-i\beta _{\mathrm{i}}}\left( \begin{array}{c}
\tilde{n}_{\mathrm{out}1}\\
\tilde{n}_{\mathrm{out}2}
\end{array}\right) \Biggr] \, .
\eea
This noise is not of the magnitude that Eqs.~(\ref{ninnout}) 
would appear to indicate.
Instead of having a coefficient of 
\( \sim \sqrt{\varepsilon _{\mathrm{RSE}}/T_{\mathrm{i}}} \), 
it has a much smaller value when \( \omega \lesssim \delta _{\mathrm{i}} \).  
The reason is that the two vacuum
fluctuations traveling in opposite directions are anticorrelated
and largely cancel each other, since they are summed in the outgoing
field \( D \).  This cancellation becomes less perfect as \( \omega  \)
grows and becomes much larger than \( \delta _{\mathrm{i}} \).  This effect is
shown in Fig.~\ref{lossfactorplot}.  

For the RSE contribution to the radiation-pressure noise, 
we are interested in how much 
the two noise fields $\tilde{\mathfrak N}_{\rm IN}$ and $\tilde{\mathfrak N}_{\rm OUT}$ 
contribute to the carrier amplitude 
fluctuation {\it at the position of the test masses}.  
Therefore, we ask what the sum of \( \tilde{\mathfrak N}_{\mathrm{IN}} \)
and \( \tilde{\mathfrak N}_{\mathrm{OUT}} \) is when they combine at the end mirrors
of the arm cavities.  Since $\tilde{\mathfrak N}_{\mathrm{OUT}}$ is superposed 
on $D$, \( \tilde{\mathfrak N}_{\mathrm{OUT}} \) must be propagated 
through the sloshing cavity and back to the arm cavity, where it is
combined with \( \tilde{\mathfrak N}_{\mathrm{IN}} \).  There is
a phase factor of \( e^{i\omega L/c} \) due to the propagation
from the internal mirror to the end mirror (in addition to the
phases acquired on the way to and inside the sloshing cavity; these
are explained below), producing 
\begin{eqnarray}
B & \rightarrow  & 
	B+e^{i\omega L/c}
	\left[ \tilde{\mathfrak N}_{\mathrm{IN}}
		-\tilde{\mathfrak N}_{\mathrm{OUT}}(1-T_{\mathrm{o}})
		\frac{e^{2i\beta _{\mathrm{s}}}}{1-T_{\mathrm{o}}
		e^{2i\beta _{\mathrm{s}}}}\right] \nonumber \\
 & \approx  & B+2T_{\mathrm{o}}
 	\sqrt{\frac{\varepsilon _{\mathrm{RSE}}}{\mathrm{T}_{\mathrm{i}}}}
 	\Biggl[ \frac{\omega (\delta _{\mathrm{i}}+\delta )
 		+i\Omega ^{2}}{\omega \delta }\tilde{n}_{\mathrm{in}}
\nonumber \\
 	&& \qquad \qquad \qquad \qquad 
 	+\frac{\omega (\delta _{\mathrm{i}}-\delta )-i\Omega ^{2}}
 		{\omega \delta }\tilde{n}_{\mathrm{out}}\Biggr] \, .
\label{ampflucRSE} 
\end{eqnarray}
where $\beta_{\rm s} = \arctan (2\omega/\delta_{\rm s})$ is the phase 
associated with the sloshing cavity.
Propagating the new $B$ to the output produces a radiation-pressure contribution
\bea
\label{NradpresRSE}
{\mathbf N}^{\mathrm{rad}\, \mathrm{pres}}_{\mathrm{RSE}} &=&
	\sqrt{\frac{\varepsilon_{\mathrm{RSE}}T_{\mathrm{o}}}
		{\mathrm{T}_{\mathrm{i}}}}e^{2i\psi }
	\left( \begin{array}{cc}
0 & 0\\
-\kappa ^{*} & 0
\end{array}\right) \nonumber \\
&& \times
\Biggl[ \frac{\omega (\delta _{\mathrm{i}}+\delta )+i\Omega ^{2}}{\omega \delta }
\left( \begin{array}{c}
\tilde{n}_{\mathrm{in}1}\\
\tilde{n}_{\mathrm{in}2}
\end{array}\right) 
\nonumber \\
&& \qquad 
+\frac{\omega (\delta _{\mathrm{i}}-\delta )-i\Omega ^{2}}{\omega \delta }
\left( \begin{array}{c}
\tilde{n}_{\mathrm{out}1}\\
\tilde{n}_{\mathrm{out}2}
\end{array}\right) \Biggr] \,.
\eea
As before, this noise does not have a magnitude 
\( \sim \sqrt{\varepsilon _{\mathrm{RSE}}/T_{\mathrm{i}}} \); it is much
smaller.  The reason is that 
when \( \tilde{\mathfrak N}_{\mathrm{OUT}} \)
travels to the sloshing cavity and back to the arms, it gains two phase 
shifts.  First is a constant phase shift of \( \pi  \),
due to the distance it traveled (twice) between the RSE and sloshing 
mirror.  The other is from the sloshing cavity, where
for frequencies much larger than the bandwidth $\delta_{\rm s}$ 
of the sloshing cavity, this phase shift is roughly \( \pi \). 
Adding these two phase shifts,
\( \tilde{\mathfrak N}_{\mathrm{OUT}} \) will appear roughly unchanged when it combines
with \( \tilde{\mathfrak N}_{\mathrm{IN}} \) in the arm cavity.  Since
these two vacuum fields are anticorrelated, there is again an effective
cancellation between the two noises at frequencies above $\delta_{\rm s}$.  
This cancellation becomes less complete
at low frequencies; see Fig.~\ref{lossfactorplot}.

We assume the fractional loss $\varepsilon_{\rm RSE} \sim 2 \times 10^{-5}$, 
since it arises primarily from losses in the RSE cavity's optical elements
(mirrors and beam splitter). (See
Appendix~\ref{app:modemismatch} for a discussion of the noise due to
mode mismatching, which we do not consider here.)

\subsection{Detection and Filter Cavities}
\label{sec:lossyfilters}

%The optical losses associated with the filter cavities have two
%origins: (i) the loss in  the optical train directing the output light
%into the filters, and (ii) the loss in the filters themselves. 
%The loss in optical train has no frequency dependence and is 
%modeled by Eq.~(\ref{includeloss}) with 
%${\cal E}_{\rm OPC} = \varepsilon_{\rm OPC}$.

First, we consider the losses involved in the detection of the 
signal (without filter cavities).  Two important
sources of photon loss are mode mismatching
associated with the local oscillator used for 
frequency-independent homodyne detection ($\varepsilon_{\rm lo}$)
and the inefficiency of the photodiode ($\varepsilon_{\rm pd}$).  
In a squeezed-input speed meter, there will
also be a circulator (with fractional loss $\varepsilon_{\rm circ}$) 
through which the squeezed vacuum is fed into the system and
through which the output light will have to pass.
These losses have no frequency dependence, so they are 
modeled by an equation of the form of
[Eq.~(\ref{includeloss})] with 
\beq
{\cal E}_{\rm OPC} (\omega)=\varepsilon_{\rm OPC} 
= \varepsilon_{\rm lo}+\varepsilon_{\rm pd}+\varepsilon_{\rm circ} 
%\sim 0.003 
\eeq
[cf. KLMTV Eq.~(104)].  The contribution to the
noise is then
\beq
{\mathbf N}_{\rm OPC} = \sqrt{\varepsilon_{\rm OPC}}
\left( \begin{array}{c}
\tilde{n}_{\mathrm{OPC}1}\\
\tilde{n}_{\mathrm{OPC}2}
\end{array}\right) \;,
\eeq
where the $\tilde{n}_{\mathrm{OPC}j}$ are linear combinations of the 
individual (independent) vacuum fields 
entering at each location (so the spectral densities of these fields are
unity and there are no cross-correlations) and propagated to the
output port.
KLMTV assumed that each of these losses is about 0.001, giving 
$\varepsilon_{\rm OPC}\sim 0.003 $.

%\beq
%\varepsilon_{\rm OPC} 
%= \varepsilon_{\rm lo}+\varepsilon_{\rm pd}+\varepsilon_{\rm circ}
%\sim 0.003 
%\eeq

We next turn our attention to optical filters on the output 
(as in the case of frequency-dependent 
homodyne detection for a squeezed-variational 
speed meter, discussed in Sec.~\ref{sec:sqvacFDdetect}).
Such cavities will have losses that
may contribute significantly to the noises of QND interferometers, as
has been seen in KLMTV.  In their Sec.~VI, KLMTV carried out a detailed analyses of
such losses; our investigation is essentially the same as theirs. 

The loss in the optical filters can come from scattering or
absorption in the cavity mirrors, which can be modeled by attributing
a finite transmissivity  $T_{\rm e}$ to the end mirrors, as we did
for the arm cavities.
The effect of lossy filters is again analogous to [Eq.~(\ref{includeloss})].
This time the loss coefficient $\cal E_{\rm F} (\omega)$ does have some frequency
dependence:
\beq
{\cal E}_{\rm F} = 2 \varepsilon_{\rm mm}+ \sum_{J=\rm I, II} \bar{\cal E}_J
	=2 \varepsilon_{\rm mm}+\frac{1}{2}\sum_{J=\rm I, II}
\left({\cal E}_{J+}+{\cal E}_{J-}\right)\,,
\label{Ef}
\eeq
where $\varepsilon_{\rm mm} \sim 0.001$ is the mode-mismatching into each 
filter cavity and where
\beq
{\cal E}_{J\pm}=\frac{4 T_e}{ T_{J}
\left[1+(\pm\omega/\delta_{J}-\xi_{J})^2\right]}
\eeq
are the loss coefficents of the two different filter cavities ($J={\rm I,II}$)
[cf.~Eqs.~(103) and (106) of KLMTV]. The noise contribution is
\beq
{\bf N}_{\rm F} = \sqrt{\cal E_{\rm F}} 
\left( \begin{array}{c}
\tilde{n}_{\rm F1} \\
\tilde{n}_{\rm F2} \\
\end{array} \right) \;.
\eeq
The weak frequency-dependence of ${\cal E}_{\rm F}$ will be neglected
(as KLMTV did), giving 
\beq
\varepsilon_{\rm F} \simeq {\cal E}_{\rm F} \sim 0.005 
\eeq
[cf.~Eqs.~(107) and (104) of KLMTV].
The value of $\varepsilon_{\rm F}$ may vary slightly for the
different optimizations we have used, but it remains less
than $0.006$.

\section[Effects due to Mode-Mismatching: A Simple Analysis]{Effects due to Mode-Mismatching: \\ A Simple Analysis}
\label{app:modemismatch}

In the practical implementation of GW interferometers,
the mismatching of spatial modes between different 
optical cavities will degrade the
sensitivity because signal power will be lost into higher-order modes and,
correspondingly, vacuum noises from those
modes will be introduced to the signal. 
In a way, this is similar to other sources of optical
loss discussed in the previous appendix. However, the higher-order modes 
do not simply get dissipated --- they too
will propagate inside the interferometer (although with a different
propagation law).  As a consequence, the exchange of energy between
fundamental and higher modes due to mode-mismatching is
\emph{coherent}, and the formalism we have been using for the loss
does not apply. In this section, we shall extend our formalism to
include one higher-order mode and give an extremely simplified model of the
mode-mismatching effects\footnote{This way of modeling the mode-mismatching
effects was suggested to us by Stan Whitcomb.}.

In a conventional interferometer (LIGO-I), the mode-mismatching comes
predominantly from the mismatch of the mirror shapes between the two arms, 
which makes
the wavefronts from the two arms different at the beam splitter. In
particular, the cancellation of the carrier light at the dark port
is no longer perfect, and additional (bright-port) noises are
introduced into the dark-port output. For our speed meter, a third
cavity---the sloshing cavity---has to be matched to the two arm 
cavities, further complicating the problem. 

In order to simplify the situation, we approximate all the waves
propagating in the corner station (the region near the beam splitter,
where the distances are short enough that ) as following the same
phase-propagation law 
as a plane wave.  The only possible source of mismatch is assumed to
come from the difference of wavefront shapes (to first order in the
fractional difference of the radii of
curvature) and waist sizes for the light beams emerging from the two
arm cavities and the sloshing cavity. Suppose, in the region of the 
corner station, we have a fiducial fundamental
Gaussian mode $\Psi^{(0)}$ (which is being pumped by the carrier) 
with waist size $w_0$ and wavefront
curvature 
$\alpha_0\equiv1/R_0$ 
that is roughly the same as 
those of the three cavities\footnote{We have chosen to use the 
curvature instead of the radius
of curvature because in this region the wavefronts are very flat.}: 
\beq
\Psi^{(0)}(x,y)\propto \frac{1}{w_0}
\exp\left(-\frac{\rho^2}{w_0^2}
	+ik\frac{\alpha_0\rho^2}{2}\right)\,,
	\; \rho=\sqrt{x^2+y^2}\,.
\eeq
At leading order in the mismatches, the fundamental modes of the three 
cavities (in the region of the corner station), which have waist 
sizes $w_{\rm J}$ and curvatures $\alpha_{\rm J}\equiv1/R_{\rm J}$ 
[${\rm J}=$n, e, or slosh (for the north arm, east arm, and sloshing cavity, 
respectively)],
can be written in the form:
\bea
&& \Psi_{\rm fnd}^{\rm J}(x,y) \nonumber \\
&& \; \propto 
	\frac{1}{w_0} \, 
	\exp \left( i k w_0^2 \, \frac{\alpha_{\rm J} - \alpha_0}{4} \right)
	\exp\left(-\frac{\rho^2}{w_0^2}+ik\frac{\alpha_0\rho^2}{2}\right)
	\nonumber \\
	&& \quad \times
	\Biggl\{ 1
	 +\left(\frac{w_{\rm J}-w_0}{4 w_0} 
		+ik w_0^2 \frac{\alpha_{\rm J}-\alpha_{\rm 0}}{16}\right)
	\nonumber \\
	&& \qquad \qquad \times
	\left[ H_2 \left( \frac{\sqrt{2}x}{w_0} \right) 
		+ H_2 \left( \frac{\sqrt{2}y}{w_0} \right) \right] \Biggr\}\,,
\eea
where $H_2 (u)$ is the second-order Hermite polynomial of $u$.
This $\Psi_{\rm fnd}^{\rm J}(\pi, y)$ can be expressed as $\Psi^{(0)}$
plus a small admixture of a higher-order mode $\Psi^{(1)}$, which consists
of equal amounts of ${\rm TEM_{02}}$ and ${\rm TEM_{20}}$
modes [and thus is orthogonal to $\Psi^{(0)}$].  This admixture
changes the waist size from $\omega_0$ to $\omega_{\rm J}$ and the curvature 
from $\alpha_0$ to $\alpha_{\rm J}$.  We can choose our 
fiducial fundamental mode $\Psi^{(0)}$ in such a
way that the two arm cavities have an opposite mismatch with it, i.e.,
$\alpha_{\rm n}+\alpha_{\rm e}=2 \alpha_0$, $w_{\rm n}+w_{\rm e}=2w_0$, and
at leading order,
\beq
\left(
\begin{array}{c}
\Psi_{\rm fnd}^{\rm n,\,e} \\
\Psi_{\rm exc}^{\rm n,\,e}
\end{array}
\right)
=
\left(
\begin{array}{cc}
1 &   \pm\mu_{\rm arm} \\
\mp\mu_{\rm arm}^* & 1
\end{array}
\right)
\left(
\begin{array}{c}
\Psi^{(0)} \\ 
\Psi^{(1)}
\end{array}
\right)\,,
\eeq
where ``exc" denotes the excited mode and the admixing 
amplitude $\mu_{\rm arm}$ is, in general, complex. We also denote the
fundamental and excited modes of the sloshing 
cavity as
\beq
\left(
\begin{array}{c}
\Psi_{\rm fnd}^{\rm slosh} \\
\Psi_{\rm exc}^{\rm slosh}
\end{array}
\right)
=
\left(
\begin{array}{cc}
1 &   \mu_{\rm slosh} \\
-\mu_{\rm slosh}^* & 1
\end{array}
\right)
\left(
\begin{array}{c}
\Psi^{(0)} \\ 
\Psi^{(1)}
\end{array}
\right)\,;
\eeq
again, $\mu_{\rm slosh}$ can be complex.
We shall also assume that the higher-order modes
involved here are far from resonance inside the cavities and
will be rejected by them, gaining a phase of $\pi$ upon reflection
from each cavity's input mirror. In the output, 
we assume the mode $\Psi^{(0)}$ is selected for
detection.  (The local oscillator associated with
the homodyne detection is chosen to have the same spatial mode as $\Psi^{(0)}$,
thereby ``selecting" $\Psi^{(0)}$.  Note that the potential mode-mismatch
effect here is already taken into account in the fractional
loss $\varepsilon_{\rm lo}$ of the local oscillator, as described in 
Sec.~\ref{sec:lossyfilters}.)

Quite naturally, we have to introduce two sets of quadrature operators to describe
the two modes. For example, for the field $P(\zeta)$ entering through the extraction
mirror, we have
\beq
{\bf \tilde{p}}^{(0)}\equiv
\left(
\begin{array}{c}
\tilde{p}_1^{(0)} \\
\tilde{p}_2^{(0)}
\end{array}
\right)\,,\quad
{\bf \tilde{p}}^{(1)}\equiv
\left(
\begin{array}{c}
\tilde{p}_1^{(1)} \\
\tilde{p}_2^{(1)}
\end{array}
\right)\,.
\eeq
For each of the three cavities, we have to decompose the optical field
into its own fundamental and excited modes, propagate them separately
and then combine them. The input--output ($a$--$b$) relation of one of the 
cavities with mirrors held
fixed can be written as
\beq
\left(
\begin{array}{c}
{\bf \tilde{b}^{(0)}} \\
{\bf \tilde{b}^{(1)}}
\end{array}
\right)
=
\left[
e^{i\Phi_{\rm fnd}}
{\bf P}_{\rm fnd}
+
e^{i\Phi_{\rm exc}}
{\bf P}_{\rm exc}
\right]
\left(
\begin{array}{c}
{\bf \tilde{a}^{(0)}} \\
{\bf \tilde{a}^{(1)}}
\end{array}
\right)\,,
\label{inoutcavityMM}
\eeq
where
\begin{mathletters}
\bea
{\bf P}_{\rm fnd} &=&
\left(
\begin{array}{c}
1 \\
\mu
\end{array}
\right)
\left(
\begin{array}{cc}
1  & 
\mu^*
\end{array}
\right)\,,  \\
{\bf P}_{\rm exc}
&=&
\left(
\begin{array}{c}
-\mu^* \\
1
\end{array}
\right)
\left(
\begin{array}{cc}
-\mu & 1
\end{array}
\right)\,,
\eea
\end{mathletters}
are the projection operators, and
$\Phi_{\rm fnd}$ and $\Phi_{\rm exc}=\pi$ are the phases gained by
the fundamental mode and excited mode after being reflected back by the
cavity. 

The mode-mismatching can cause both shot and radiation pressure noises
at the output, giving:
\beq
{\bf \tilde{q}^{(0)}}
\rightarrow 
{\bf \tilde{q}^{(0)}}+{\bf N}^{\rm shot}_{\rm MM}+{\bf N}^{\rm rad\,pres}_{\rm MM}\,.
\eeq
Assuming the mirrors are held fixed and applying the new input--output relations (\ref{inoutcavityMM}) of the
non-perfect cavities, we get the following shot noise in the output
(to leading order in $\mu_{\rm arm}$ and $\mu_{\rm slosh}$):
\bea
{\bf N}_{\rm MM}^{\rm shot} 
&=&
-e^{i\psi}{\mu_{\rm arm}^*}
\sqrt{\frac{4}{T_{\rm o}}}
\frac{\sqrt{T_{\rm p}}}{1+\sqrt{1-T_{\rm p}}}\nonumber \\
&& \qquad \qquad \times
\frac{1-\sqrt{1-T_{\rm i}}}{\sqrt{T_{\rm i}}} 
\frac{\omega\delta}{|{\cal L}(\omega)|} \,
{\bf \tilde{i}^{(1)}} \nonumber \\
&\approx&
e^{-i\psi}\mu^*_{\rm arm}\sqrt{\frac{T_{\rm i}T_{\rm
      p}}{4T_{\rm o}}}\frac{\omega\delta}{|{\cal L}(\omega)|} 
{\bf \tilde{i}^{(1)}}\,;
\label{MMshot}
\eea
see Eq.~(\ref{lossform}). The quantity ${\bf \tilde{i}^{(1)}}$ 
refers to the excited mode of the noise
coming in the bright port [$I(\zeta)$ in Fig.~\ref{fig:oneport3IM}].

The main results embedded in Eq.~(\ref{MMshot}) are
\begin{itemize}
\item[(i)] the mode-mismatching with the sloshing cavity does not 
give any contribution at leading
order in $\mu$, and
\item[(ii)] the mode-mismatching shot noise comes from the higher-order
mode entering from the {\it bright port}, strongly suppressed by the
presence of the internal and power-recycling mirrors.
\end{itemize}
These two effects are both due to the coherent interaction between the
fundamental ($\Psi^{(0)}$) and excited  ($\Psi^{(1)}$) modes (of our
idealized cavity), in which
energy is not simply dissipated from $\Psi^{(0)}$ but exchanged
coherently between the two modes as the light flows back and forth
between the sloshing cavity and the arm cavities. Detecting an 
appropriate linear combination of
the two modes can then be expected to reverse the effect of mode mismatching. 
In our case, the properties of the cavities are carefully chosen 
such that $\Psi^{(0)}$ itself
is the desired detection mode (for the sloshing mismatch).  Consequently,
the mode mismatching with the sloshing cavity does not contribute at 
leading order [item (i) above].  
Regarding item (ii), the mismatch of the two arm cavities does give
rise to an additional noise, but it can only come from the higher mode
in the bright port, because at leading order in mismatches, (a) the
propagation of $\Psi^{(0)}$ from the 
bright port to the dark port is suppressed and (b) there is no
propagation of dark-port $\Psi^{(1)}$ into dark-port $\Psi^{(0)}$
since we have chosen $\Psi^{(0)}$ in such a way that the 
two arm cavities have exactly opposite mismatches with it.
  
The reason why
this noise is suppressed by the factor $1/T_{\rm p}$ is simple: because
$\Psi^{(1)}$ is not on resonance with the composite cavity formed by
the power-recycling mirror and the arm cavities, its fluctuations
inside the system (like its classical component) are naturally 
suppressed by a factor
$1/\sqrt{T_{\rm p}}$ compared to the level outside the cavity. 
%[This means that, if the laser
%is shot-noise limited in the excited mode, then the fluctuations of
%the in-cavity field will be significantly less than the shot noise in
%this mode.] 
The
reason for the factor of $1/T_{\rm i}$ is similar: the $\Psi^{(1)}$
mode does not resonate within the system formed by the arm cavities
and the RSE mirror and will consequently be suppressed.

By computing at the fields at the end mirrors and from them the fluctating
radiation pressure, we obtain the
radiation-pressure noise due to mode-mismatching:
\beq
{\bf N}_{\rm MM}^{\rm rad\,pres}=-\frac{e^{2 i\psi}}{2}\mu_{\rm arm}^*\sqrt{\frac{T_{\rm i} T_{\rm p}}{4T_{\rm o}}}
\left(
\begin{array}{cc}
0 & 0 \\
-\kappa^* & 0
\end{array}
\right)
{\bf \tilde{i}^{(1)}}\,.
\label{MMrad}
\eeq
This radiation-pressure noise is 
suppressed by a factor similar to the shot noise.  

By comparing Eqs.~(\ref{MMshot}) and (\ref{MMrad}) 
with, e.g., Eqs.~(\ref{NAESloss}), we see that 
mode mismatching produces noise with  essentially the same 
form as optical-element losses from the
arms, extraction mirror and sloshing cavity (AES), with (assuming the
input laser is shot-noise limited in the higher modes)
\beq
\varepsilon_{\rm MM}=\frac{T_{\rm i}T_{\rm p}}{4}|\mu_{\rm arm}^*|^2\,.
\eeq
The factor $T_{\rm i}T_{\rm P}/4$ happens to be the ratio
between the input power (at the power-recycling mirror) and the
circulating power, which will be $\sim 10^{-4}$. Suppose 
$\Re(\mu_{\rm arm})\sim\Im(\mu_{\rm arm})\sim0.03$.  The effect 
of mode-mismatching will then be much
less significant ({\it in our simple model}) than the losses from the 
optical elements. 

It should be evident that other imperfections in the cavity mirrors,
which cause admixtures of other higher-order (``excited") modes,
will lead to similar ``dissipation factors," 
${\cal E}_{\rm MM} \sim \frac{T_{\rm i} T_{\rm p}}{4} |\mu^*_{\rm arm}|^2$.
For this reason, we expect mode mismatching to contribute negligibly to the 
noise, and we ignore it in the body of the paper.

\section{Transmissivity Mismatch between the Internal Mirror and the RSE Mirror}
\label{sec:mirrormismatch}

Recall from Sec.~\ref{sec:Introduction} that when the internal and RSE mirrors
have the same transmissivity, their effects on the gravity-wave sideband
cancel.  If, however, 
the transmissivity of the internal mirror, \( T_{\mathrm{i}} \),
is not perfectly matched by that of the RSE mirror, \( T_{\mathrm{RSE}} \),
then this cancellation will no longer be perfect.  As a result, the
RSE cavity (i.e., the cavity between the internal and RSE mirrors) will 
have the same effect as an additional mirror (with a small reflectivity).
Suppose the transmissivity of this effective mirror is 
\( T_{\mathrm{RSE}}=(1+\varepsilon _{\mathrm{RSE}})T_{\mathrm{i}} \).  Then
a simple calculation yields its (amplitude) reflectivity:\begin{equation}
\label{mismatch}
\mu =\frac{\sqrt{1-T_{\mathrm{i}}}-\sqrt{1-T_{\mathrm{RSE}}}}
	{1-\sqrt{1-T_{\mathrm{i}}}\sqrt{1-T_{\mathrm{RSE}}}}
	\approx \frac{\varepsilon _{\mathrm{RSE}}}{2\sqrt{1-T_{\mathrm{i}}}}
	\approx \frac{\varepsilon _{\mathrm{RSE}}}{2}\, .
\end{equation}

Adding this effective mirror with reflectivity $\mu$ to our interferometer yields
a new set of input--output relations similar to Eq.~(\ref{qsoutKimbleform}),
but with modified \( \kappa  \) and \( \psi  \). The functional form
of \( \kappa  \) can be maintained by appropriately redefining the quantities
\( \Omega  \) and \( \delta  \).  To leading order
in \( \mu  \), we obtain
\begin{equation}
\label{kappamismatch}
\kappa \rightarrow \kappa _{\mathrm{TM}}
=\frac{\Omega _{\mathrm{I}}^{3}\delta _{\mathrm{TM}}}
{(\omega ^{2}-\Omega _{\mathrm{TM}}^{2})^{2}+\omega ^{2}\delta ^{2}_{\mathrm{TM}}}\,,
\end{equation}
with
\begin{equation}
\label{redefmismatch}
\Omega \rightarrow \Omega _{\mathrm{TM}}=(1-\mu )\Omega \, ,\quad 
\delta \rightarrow \delta _{\mathrm{TM}}=(1-2\mu )\delta \, .
\end{equation}
Consequently, we can re-optimize the system to compensate for this
transmissivity-mismatch effect.

\end{document}